\documentclass[1p]{elsarticle}
\usepackage{graphicx}
\usepackage{amsmath}
\usepackage{amsfonts}
\usepackage{amssymb}
\usepackage{ulem}
\usepackage{physics}
\usepackage{xcolor}
\graphicspath{{./Figures/}}
\usepackage{float}
\usepackage{mathrsfs}

\def\bra#1{{\left\langle#1\right\vert^{}}}

\def\ket#1{{\left\vert#1\right\rangle}}
\def\abs#1{{\left|#1\right|}}

\def\tFWHM{\tau_\text{FWHM}}
\def\tctrl{\tau^\text{ctrl}_\text{FWHM}}
\def\Deltau{\Delta\tau^\text{ctrl}}

\definecolor{darkgreen}{RGB}{34,139,34}

\begin{document}

\title{Broadband Quantum Memory in Atomic Ensembles}
\author[1,2]{Kai Shinbrough\footnote{kais@illinois.edu}$^,$}
\author[1,2]{Donny R. Pearson Jr.}
\author[3]{Bin Fang}
\author[1,2]{Elizabeth A. Goldschmidt}
\author[1,2]{Virginia O. Lorenz}

\address[1]{Department of Physics, University of Illinois Urbana-Champaign, 1110 West Green Street, Urbana, IL 61801, USA}
\address[2]{Illinois Quantum Information Science and Technology (IQUIST) Center, University of Illinois Urbana-Champaign, 1101 West Springfield Avenue, Urbana, IL 61801, USA}
\address[3]{Center for Dynamics and Control of Materials, The University of Texas at Austin, Austin, TX 78712, USA}

\date{\today}

\begin{abstract}
    Broadband quantum memory is critical to enabling the operation of emerging photonic quantum technology at high speeds. Here we review a central challenge to achieving broadband quantum memory in atomic ensembles---what we call the `linewidth-bandwidth mismatch' problem---and the relative merits of various memory protocols and hardware used for accomplishing this task. We also review the theory underlying atomic ensemble quantum memory and its extensions to optimizing memory efficiency and characterizing memory sensitivity. Finally, we examine the state-of-the-art performance of broadband atomic ensemble quantum memories with respect to three key metrics: efficiency, memory lifetime, and noise.  
\end{abstract}

\maketitle

\tableofcontents

\section{Introduction}

\subsection{Motivation}

The fundamental challenges behind distributing quantum information using light are light's fixed speed and ubiquitous propagation losses in optical fiber. Given the inability to copy quantum states \cite{wootters1982single}, this means that, in order to engage in quantum information processing using systems in separate locations, one must find a way to store the quantum information contained in an optical field for a time commensurate with the travel time between locations. This holds for long-distance operations where ms-scale storage is required for round-trip communication \cite{sangouard2011quantum,muralidharan2016optimal} as well as on-chip or integrated devices where ns-scale delays allow more complex quantum logic \cite{kitching2018chip,lu2021advances,bremer2020cesium,maisch2020controllable,siverns2019demonstration,akopian2011hybrid}. Such a quantum memory must store an incoming photonic quantum state, with quantum information encoded in one (or more) of its degrees of freedom, and faithfully retrieve that photonic state without altering its quantum information or adding noise. The most intuitive application of this primitive operation relates to photon synchronization: 
two photons arriving at, for example, a node in a quantum network \cite{kimble2008quantum}, may have traveled long and disparate distances to the node and require some form of quantum memory if they are both to impinge on a beamsplitter at the same time and affect a Bell state measurement \cite{tittel1998violation}, entanglement swapping \cite{pan1998experimental}, quantum teleportation \cite{bouwmeester1997experimental}, or almost any other quantum networking protocol. This is perhaps the most straightforward application of quantum memory; however, quantum memory is also of critical importance for all-optical quantum computing \cite{kok2007linear}, quantum communication \cite{shor2000simple}, enhanced measurement and sensing \cite{gottesman2012longer}, and local quantum gates \cite{campbell2014configurable}.

There are many potential architectures for implementing quantum memory effectively. Here we focus on the storage of quantum states of light in collective states of atomic ensembles. This general class of schemes is widely applicable across different wavelength and bandwidth regimes, and is only fundamentally limited by the coherence and optical depth of the atomic system, which can be chosen or engineered to be suitable for applications, albeit typically at the expense of other important parameters. In this work, we focus specifically on implementations of atomic ensemble quantum memory in the broadband regime, which we consider to be photon bandwidths greater than 10 MHz. Broadband memory operation is of unique importance for the implementation of quantum photonic applications at high speeds, as high clock rates and processing speeds imply the use of photons that are short in duration and therefore broad in bandwidth.

This chapter concerns the use of both atomic or atom-like ensemble systems for broadband optical quantum memory. Atom-like systems include rare-earth ions doped in solids, molecular gases, phonons in solids, and any platform in which there exists an ensemble of particles with at least three energy levels on which the memory interaction can be based. For ease of notation we refer to both atomic and atom-like systems as ``atomic.'' Note that in this work we do not consider ``atomic-ensemble quantum memories'' to include the creation of single photons entangled with long-lived matter excitations \cite{duan2001long}, long-lived excitations in single atoms or ions \cite{kielpinski2001decoherence, wang2017single}, or other means of transducing photonic quantum information into a material platform \cite{bhaskar2020experimental}. For more information on these alternative mechanisms for engineered atom-photon interactions, we refer the reader to the reviews of Refs.~\cite{sangouard2011quantum,reiserer2015cavity}.

Even within the relatively narrow scope of broadband quantum memory using collective atomic states, there exist many physically distinct quantum memory protocols and many distinct hardware implementations. Each protocol and hardware implementation possesses particular advantages and disadvantages. We provide some context for these relative advantages and disadvantages by first introducing the fundamentals and the merits of atomic ensembles as quantum memories in Sec.~\ref{ensemble_sec}. We then discuss a critical problem for atomic ensemble quantum memory in Sec.~\ref{bwlw_sec}, the linewidth-bandwidth mismatch problem. We present the metrics used to quantify memory performance in Sec.~\ref{metric_sec}, before launching into a comprehensive discussion of atomic memory protocols and hardware implementations and their respective advantages in Sec.~\ref{ProtocolHardware_Sec}. In Sec.~\ref{theory_sec}, we review the theory of atomic ensemble quantum memory, including the various forms of the equations of motion in the presence of homogeneous and inhomogeneous broadening and the mathematical approximations that lead to each physical protocol. With these theoretical foundations in mind, we then discuss the theoretical tools developed for the optimization of memory efficiency and the recently developed theoretical tools for investigating memory sensitivity \cite{shinbrough2022variance}, which describes the behavior of atomic ensemble quantum memory in the presence of experimental fluctuations and drift. Finally, in Sec.~\ref{SotA_sec}, we turn to the state-of-the-art performance of broadband atomic-ensemble quantum memories in the literature, providing empirical evidence for the advantages and disadvantages discussed in Sec.~\ref{ProtocolHardware_Sec}. We focus on three metrics of particular importance to broadband atomic ensemble quantum memory: efficiency (Sec.~\ref{eff_sec}), memory lifetime (Sec.~\ref{lifetime_sec}), and noise (Sec.~\ref{noise_sec}).

\subsection{Ensemble Atomic Systems}\label{ensemble_sec}

\begin{figure}[t]
	\centering
	\includegraphics[width=0.75\linewidth]{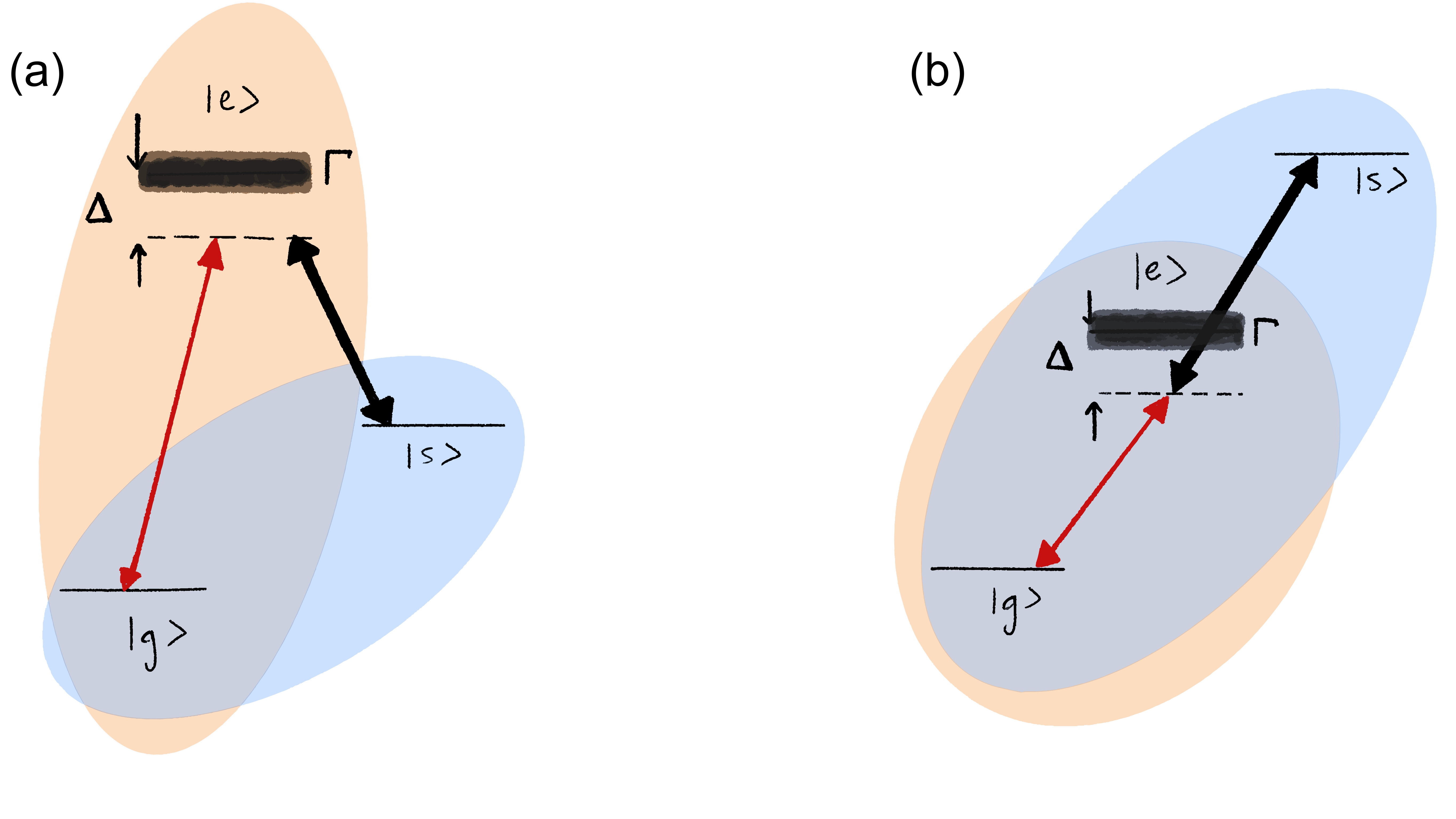}
	\vspace{-1em}
	\caption{Schematic of the atomic level structures for (a) $\Lambda$-type and (b) ladder-type ensemble quantum memory. Each structure possesses a ground state $\ket{g}$, intermediate excited state $\ket{e}$ (with population decay rate $\Gamma=2\gamma$), and metastable storage state $\ket{s}$. The signal field (red) and control field (black) are typically kept in two-photon resonance with detuning $\Delta$ from the excited state. Orange and blue shaded regions correspond to atomic polarization and spin wave coherences, respectively.}
	\label{fig_levelstruc}
\end{figure}

 Ensemble atomic systems are well-suited for quantum memory as they possess high optical depths, controllable frequency (based in part on atomic species, in part on the chosen detuning from atomic transitions), long-lived atomic superposition states, low sensitivity to experimental noise, and arbitrary storage time. As discussed below, however, atomic systems often suffer from linewidth-bandwidth mismatch, noise from the necessary optical control fields, and undesirable broadening mechanisms. Different atomic level structures have been employed experimentally for atomic ensemble quantum memory, including ladder-type \cite{finkelstein2018fast,kaczmarek2018high,thomas2022single,finkelstein2021continuous,davidson2022fast}, M-type \cite{qiu2019coherent,pengbo2006dark}, and others \cite{ham2018a,hetet2008electro,wei2020memory}; however, the atomic $\Lambda$-type level structure is the most common, and all level structures typically obey the same underlying atom-photon interaction physics (see Fig.~\ref{fig_levelstruc} for the $\Lambda$-type and ladder-type structures).  

In the memory interaction, the `signal' field of interest, which possesses some quantum information encoded in one of its degrees of freedom, is tuned on, near, or off resonance with the ground-to-excited-state transition ($\ket{g}\leftrightarrow\ket{e}$ in Fig.~\ref{fig_levelstruc}) of the atoms. We note that this signal field may be a single photon, or a more general photonic quantum state, with either discrete or continuous variable quantum information encoded in its degrees of freedom (though typically more care must be taken when using continuous variables \cite{jensen2011quantum}). For the remainder of this chapter, we typically focus on the case of single-photon signal fields. The classical `control' field, possessing many photons, is tuned to the excited-to-storage-state transition ($\ket{e}\leftrightarrow\ket{s}$). These two atomic transitions are typically assumed to be dipole allowed, and the $\ket{s}\leftrightarrow\ket{g}$ transition is assumed to be forbidden. Atoms entering the $\ket{s}$ state are thus metastable. In this typical situation, the time-domain Maxwell-Bloch equations describing the memory interaction are \cite{arecchi1965theory,gorshkov2007photon_1,gorshkov2007photon_2,gorshkov2007photon_3,gorshkov2007universal,nunn2008quantum}:

\begin{align}
    \label{Aeq_t_hom}\partial_z A(z,\tau) &= -\sqrt{d} P(z,\tau)\\
    \label{Peq_t_hom}\partial_\tau P(z,\tau) &= -\bar{\gamma} P(z,\tau) + \sqrt{d} A(z,\tau) - i\frac{\Omega(\tau)}{2} B(z,\tau)\\
    \label{Beq_t_hom}\partial_\tau B(z,\tau) &= -\gamma_B B(z,\tau) -i\frac{\Omega^*(\tau)}{2} P(z,\tau),
\end{align}

\noindent where $z$ represents the one-dimensional spatial dimension of the atomic ensemble normalized to the ensemble length [i.e., $z=0$ ($z=1$) represents the beginning (end) of the ensemble]; $\tau = t-z/c$ represents time measured in the comoving frame of the photon ($t$ represents time in the lab frame) normalized to the excited-state coherence decay rate $\gamma = \Gamma/2$ ($\Gamma$ is the total excited-state population decay rate, or the linewidth of the $\ket{g}\leftrightarrow\ket{e}$ transition); $A(z,
\tau)$ is the spatially and temporally dependent signal photon field; $P(z,\tau)$ and $B(z,\tau)$, referred to as the atomic polarization and spin wave fields, respectively, are macroscopic field operators representing the atomic coherences $\ket{g}\leftrightarrow\ket{e}$ and $\ket{g}\leftrightarrow\ket{s}$, which are delocalized across the length of the medium and are shown in Fig.~\ref{fig_levelstruc} as orange and blue shaded regions; $d$ is the resonant optical depth of the memory; $\bar{\gamma} = (\gamma-i\Delta)/\gamma$ is the normalized complex detuning, where the two-photon detuning $\Delta$ is shown schematically in Fig.~\ref{fig_levelstruc}; and $\Omega(\tau)$ is the control field Rabi frequency coupling the $\ket{e}$ and $\ket{s}$ states. All atomic population is assumed to start in the ground state, and the metastable storage state is assumed to have a coherence decay rate $\gamma_B$ that is much smaller than the excited state decay rate ($\gamma_B\ll1$, in normalized units).

By inspection of Eqs.~\eqref{Aeq_t_hom}-\eqref{Beq_t_hom}, we note that the photon field acts as a source for the atomic polarization with coupling constant $\sqrt{d}$, which then decays and accumulates temporal phase according to $\bar{\gamma}$. The atomic polarization then acts as a source for the spin wave field with coupling constant $\Omega(\tau)/2$, which then decays exponentially at a rate of $\gamma_B$. Depending on the sign of each of the fields, the roles of source and sink can be reversed, allowing for reversible mapping of population ultimately from the photon field to the spin wave and back.

Options other than atomic ensemble systems exist for optical quantum memory, including notably the use of optical delay lines, free-space optical cavities \cite{bouillard2019quantum,pittman2002single,kaneda2015time,kaneda2019high,xiong2016active,makino2016synchronization,pittman2002cyclical,pang2020hybrid}, optical fibers \cite{xiong2016active,clemmen2018all,magnoni2019performance}, and optical fiber cavities \cite{bustard2022toward}. These quantum memories take the simplest approach to photon storage by merely increasing the path length traveled by the photon instead of transducing the photon into a material excitation. As such, these delay-based optical quantum memories provide an important point of comparison for atomic ensemble quantum memories. Delay-based memories possess their own disadvantages, however, including fixed-increment delay times, unavoidable optical losses, slow switching speed ($\sim$MHz), and high sensitivity to thermal fluctuations and air currents. Nevertheless, delay-based quantum memories are in principle agnostic to photon bandwidth and exhibit some of the highest efficiencies and longest memory lifetimes in the broadband regime (see Sec.~\ref{SotA_sec}), limited only by losses in the optical path. Delay-based memories additionally typically exhibit ultra-low-noise operation. As shown in Ref.~\cite{cho2016highly}, only a few atomic ensemble quantum memories can currently outperform fiber delay lines in terms of memory efficiency and lifetime, where the total memory efficiency for fiber delay lines is taken as $\eta(\tau) = 10^{-\varepsilon c_n \tau/10}$ for a fiber with loss $\varepsilon$ (typically given in units of dB/km) and a speed of light in the fiber of $c_n$ (the necessary length of fiber is therefore $L = c_n \tau$). This calculation ignores coupling losses into and out of the fiber, but serves as an important point of comparison. 
For particularly long-lived storage, matter-based systems are likely the only option given the sub-ms limits of propagation delay techniques. In the case of fiber-based delay lines, for example, assuming state-of-the-art 0.2 dB/km loss \cite{agrawal2012fiber}, the $1/e$ lifetime of a single fiber loop memory (without switchable delay, which introduces additional loss) is only $\sim$100 ns. Similar constraints on memory lifetime exist for free-space delay line memories as well, in which case the most optimistic memory lifetime is $\sim$10 $\mu$s \cite{bouillard2019quantum,pittman2002single,kaneda2015time,kaneda2019high,xiong2016active,makino2016synchronization,pittman2002cyclical,pang2020hybrid}. 

Another important point of comparison between fiber-based and atomic-ensemble quantum memory concerns the distortion of broadband pulses of light in fiber due to group velocity dispersion. As the different frequencies of light in broadband photons propagate in fiber at different velocities, broadband photons experience temporal stretching and distortion described in the frequency domain by $A_\text{out}(\omega) = A_\text{in}(\omega)e^{i\beta(\omega-\omega_0)^2 L/2}$, where $\beta$ is the second-order group velocity dispersion of the fiber (typically given in units of fs$^2$/mm) evaluated at the center frequency $\omega_0$, and $L$ is the fiber length. By assuming an initially Gaussian input pulse with angular $1/e^2$ frequency bandwidth $\sigma_\omega = \pi BW / \sqrt{2\ln 2}$ (where $BW$ is the full-width at half-maximum bandwidth; see Sec.~\ref{metric_sec}), Fourier transforming into the temporal domain, and applying Eq.~\eqref{fidelityeq}, one can calculate the fidelity of a fiber-based memory as a function of storage time $t$ and photon bandwidth. Inverting this relationship and considering a fixed target memory fidelity $\mathcal{F}_0$, one can derive the following tradeoff between storage time and fiber memory bandwidth:

\begin{equation}
    t = \frac{\sqrt{1-\mathcal{F}_0^2}}{\mathcal{F}_0 \sigma_\omega^2 \beta c_n},\label{fiberTFBW}
\end{equation}

\noindent which demonstrates that either high-fidelity, large-bandwidth memory is possible at short storage times, or high-fidelity, long-storage-time memory is possible at narrow bandwidths, but all three (high-fidelity, large-bandwidth, long-storage-time) are not possible simultaneously in standard fiber. This tradeoff can be alleviated if an appropriate length of dispersion-compensating fiber is spliced onto the end of standard fiber; whether this is necessary or not for a given storage time, bandwidth, and target fidelity can be derived from Eq.~\eqref{fiberTFBW}.

While today ensemble-based quantum memories compare mostly poorly to delay lines in terms of efficiency, bandwidth, lifetime, and noise,  there is substantial room for improvement and reason to believe that further work will lead to important advances. In particular, the extremely high losses in on-chip waveguides make integrated photonic systems a place where ensemble memories may be able to make a major difference in the near term. In on-chip photonic devices, coupling on- and off-chip to fiber may introduce more loss than even low-efficiency, evanescently coupled atomic memory, and certainly more loss than chip-integrated rare-earth memories. Attempts to integrate atomic ensembles on chip and in fiber have met with some success to date \cite{sprague2013efficient,sprague2014broadband,peters2020single,bajcsy2009efficient,gouraud2015demonstration,heshami2016quantum,patnaik2002slow,bremer2020cesium,maisch2020controllable,siverns2019demonstration,akopian2011hybrid} and there is hope for improvement. Atomic ensemble memories also possess the capacity for some basic quantum optical processing that is absent among delay-based memories. These include the capacity for built-in mode conversion, shaping, and sorting, multiphoton quantum state preparation \cite{nunn2013enhancing,finkelstein2018fast}, local quantum processing \cite{campbell2014configurable}, arbitrary spectral-temporal and polarization mode conversion, and preparation of arbitrary superposition states that are either not possible through other means or not possible at the same rate or bandwidth \cite{reim2012multipulse,saglamyurek2018coherent,heller2022raman}. 

Mode shaping and conversion are processing tasks particularly well-researched in atomic ensemble memories. In the broadband regime, this typically focuses on shaping and conversion in the temporal and frequency domains, although polarization conversion has also been demonstrated \cite{england2015storage}. Demonstrations to date include frequency and bandwidth conversion in a Raman memory in room-temperature diamond \cite{fisher2016frequency}; frequency conversion in room-temperature molecular hydrogen \cite{bustard2017quantum}; temporal and frequency multiplexing, arbitrary temporal shaping, and temporal stretching and compression in an AFC memory in cryogenic Tm:LiNbO$_3$ \cite{saglamyurek2014integrated}; temporal beamsplitting in a warm cesium Raman memory \cite{reim2012multipulse}; temporal beamsplitting and temporal stretching and compression in a laser-cooled rubidium ATS memory \cite{saglamyurek2018coherent}; and temporal compression, stretching, and beamsplitting in a laser-cooled rubidium Raman memory \cite{heller2022raman}. Taken together, these demonstrations show the capacity for temporal, frequency, duration, and bandwidth conversion of atomic-ensemble memories, which may be of practical application in developing quantum networks and quantum information processors. We note, however, that additional work is needed to demonstrate this utility and to extend the characterization of atomic ensembles as mode converters to include phase as well as amplitude manipulation. 

A problem is posed for broadband atomic ensembles beyond roughly MHz bandwidths, which is the lack of sufficiently fast classical electronic switches to effectively make use of the broad optical bandwidths demonstrated. Fortunately, ongoing work toward GHz and THz optical switches shows promise toward alleviating this concern \cite{oza2013entanglement,purakayastha2022ultrafast,alarcon2020polarization,hall2011ultrafast,england2021perspectives,kupchak2019terahertz,volz2012ultrafast,kupchak2017time,friberg1987ultrafast,almeida2004all,nozaki2010sub,kanbara1994highly}.

\subsection{Linewidth-Bandwidth Mismatch}\label{bwlw_sec}

\begin{figure}[t]
	\centering
	\includegraphics[width=0.9\linewidth]{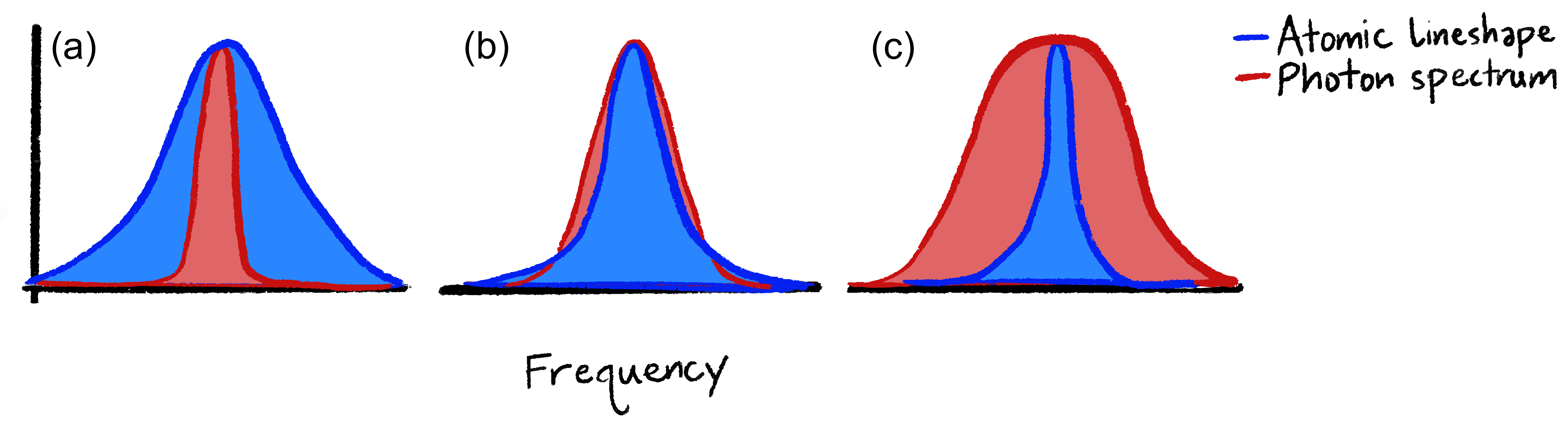}
	\vspace{-0.5em}
	\caption{Schematic of linewidth-bandwidth mismatch. In atomic ensemble quantum memories, the photon spectrum (bandwidth) may be (a) smaller than the atomic lineshape (linewidth), (b) matched, or (c) mismatched.}
	\label{fig_lwbw}
\end{figure}

Broad spectral bandwidth is of particular importance for quantum applications, and we highlight this aspect of quantum memory performance in this chapter. Quantum memory operation with broadband photons, corresponding to short photon durations, enables the use of large clock and qubit processing rates. Broadband operation presents a natural problem for standard atomic ensemble quantum memory, however, as the linewidths involved in atomic ensembles are typically narrow, and narrow storage-to-ground-state-transition linewidths are desirable for long memory lifetimes.  We refer to the problem of storing broadband photons with high efficiency in typically narrowband atomic ensemble quantum memories as the ‘linewidth-bandwidth mismatch’ problem. Figure~\ref{fig_lwbw} shows a schematic of the different situations possible for broadband photons; high memory efficiency is, in general, more easily achievable for the cases of photon bandwidths smaller than the atomic excited state linewidth [Fig.~\ref{fig_lwbw}(a)] or when the photon bandwidth is matched to the atomic linewidth [Fig.~\ref{fig_lwbw}(b)], compared to the case when the two are mismatched [Fig.~\ref{fig_lwbw}(c)]. More specifically, we consider the linewidth-bandwidth mismatch problem to encapsulate the empirical trend of decreasing memory efficiency with increasing photon bandwidth; evidence for this trend is presented in Sec.~\ref{SotA_sec}. This trend is physically a result of the difficulty of absorbing broadband photons along a narrowband transition --- if a broadband photon cannot be absorbed along the $\ket{g}\rightarrow\ket{e}$ transition, it is difficult to affect the memory operation described by the equations in Sec.~\ref{ensemble_sec}. This is easiest to see in the case of the absorb-then-transfer protocol (discussed more in Sec.~\ref{protocol_sec}), which explicitly relies on linear absorption of the signal field. The overall memory efficiency cannot be high without high-efficiency absorption.

\begin{figure}[t]
	\centering
	\includegraphics[width=1\linewidth]{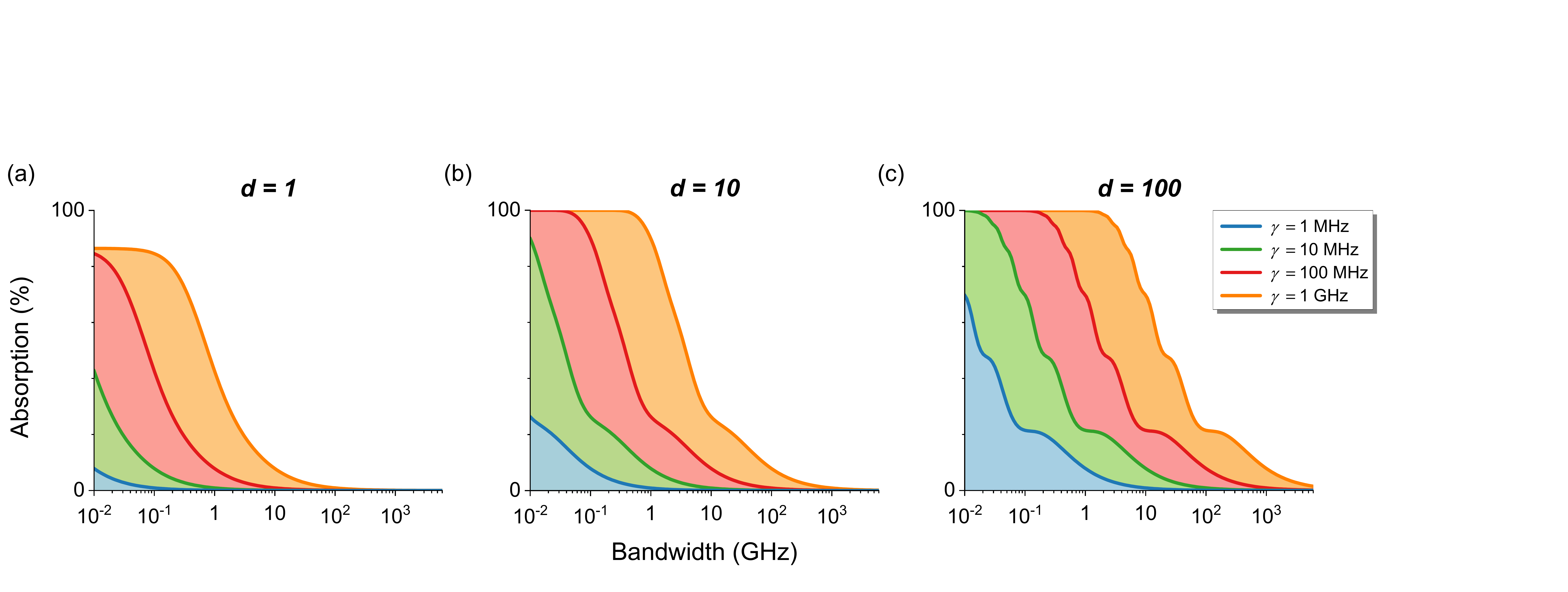}
	\vspace{-1.5em}
	\caption{Calculated upper bounds on memory efficiency based on linear absorption, demonstrating the linewidth-bandwidth mismatch problem for photon bandwidths between 10 MHz and 1 THz, assuming excited state linewidths of 1, 10, 100 MHz, and 1 GHz, and peak optical depths of (a) $d = 1$, (b) $d = 10$, and (c) $d = 100$.}
	\label{fig_lwbw_sim}
\end{figure}

We can model this phenomenon heuristically by considering the efficiency of linear absorption in a dense atomic ensemble. Linear absorption can be modeled by Eqs.~\eqref{Aeq_t_hom}-\eqref{Beq_t_hom} of Sec.~\ref{ensemble_sec} in the absence of a control field [$\Omega(\tau)=0$]:

\begin{align}
    \partial_z A(z,\tau) &= -\sqrt{d} P(z,\tau)\\
    \partial_\tau P(z,\tau) &= -\bar{\gamma} P(z,\tau) + \sqrt{d} A(z,\tau),
\end{align}

\noindent where we define absorption as $\int_{-\infty}^{\infty}d\tau \abs{A_\text{out}(\tau)}^2 \big/ \abs{A_\text{in}(\tau)}^2$ for a Gaussian-shaped input temporal distribution $A_\text{in}(\tau)$ and an output distribution $A_\text{out}(\tau) = A(\tau,z=1)$. This absorption is plotted in Fig.~\ref{fig_lwbw_sim} for photon bandwidths between 10 MHz and 1 THz, assuming excited state linewidths of 1, 10, 100 MHz, and 1 GHz. Figure \ref{fig_lwbw_sim}(a)-(c) show absorption for optical depths of $d=1$, 10, 100, respectively. In general, the larger excited state linewidths are able to absorb broadband photons with higher efficiency as they have lesser linewidth-bandwidth mismatch. Absorption can also be increased by increasing optical depth, but this is typically a less effective strategy for increasing efficiency than increasing atomic linewidth at the same optical depth. The oscillations that appear in Fig.~\ref{fig_lwbw_sim}(b)-(c) are due to periodic backaction of the atomic ensemble onto the photon temporal distribution and the formation of zero-area photon pulses \cite{costanzo2016zero}. The takeaway from Fig.~\ref{fig_lwbw_sim} is that high efficiency at broad photon bandwidths is difficult to achieve with (relatively) narrowband atomic transitions, and this is due to linewidth-bandwidth mismatch. This issue introduces a trade-off between efficiency and lifetime, since narrowband atomic transitions are useful for preserving the coherence of collective atomic states for long timescales, but lead to lower memory efficiency for broadband pulses.

One resolution to the linewidth-bandwidth mismatch problem has been demonstrated in recent work in hot atomic barium \cite{shinbrough2022broadband,shinbrough2022efficient}. This approach relies on large, tunable modification of the homogeneous excited state linewidth to better match broadband photons, where the linewidth modification is mediated by the intentional and controllable introduction of collisional dephasing. This technique creates broad homogeneous excitation linewidths that allow for complete absorption and storage of broadband photonic quantum states in a system that would otherwise exhibit relatively narrowband transitions. This approach yields a demonstrated storage efficiency of 95.6$\pm$0.3\% for ultrashort photons (500 fs), which to our knowledge is the highest measured storage efficiency to date for any atomic-ensemble memory with bandwidth $>$10 MHz, and rivals the efficiencies demonstrated in delay-line and fiber-based memories, as is discussed further in Sec.~\ref{SotA_sec}.

The controlled homogeneous broadening approach described in Refs.~\cite{shinbrough2022broadband,shinbrough2022efficient} constitutes a novel technique for alleviating the linewidth-bandwidth mismatch problem, but it is not the only technique. In particular, the use of far-off-resonant schemes for quantum memory in atomic ensembles is well known, and similarly alleviates the linewidth-bandwidth mismatch problem, albeit through a different mechanism. In far-off-resonant memory protocols, memory bandwidth is not limited by the narrowband atomic transitions, as the large detuning adiabatically eliminates the excited state. This elimination of the excited state also eliminates the restriction on memory bandwidth posed by the excited state linewidth; however, this advantage comes at the cost of large control field power necessary to drive the off-resonant two-photon transition. This requirement of large control field powers can be difficult to satisfy in experiment, and also tends to lead to low efficiency at broad photon bandwidths (see Sec.~\ref{SotA_sec}). By contrast, the technique of Refs.~\cite{shinbrough2022broadband,shinbrough2022efficient} can be applied on or near resonance, and therefore requires significantly less control field power than the far-off-resonant schemes, and can achieve similar memory bandwidths. A third technique for alleviating linewidth-bandwidth mismatch relies on inhomogeneous broadening of the otherwise narrowband atomic transitions, either in a controlled fashion (i.e., in the CRIB protocol; see Sec.~\ref{protocol_sec}) or by the choice of hardware that intrinsically possesses inhomogeneity (i.e., rare-earth ions doped in solids; see Sec.~\ref{solids_sec}). This technique can in principle be used to create inhomogeneous linewidths that match broad photon bandwidths, and therefore allow for high-efficiency, broadband quantum memory; however, the external fields necessary in the controllable case tend to be large and experimentally challenging to generate, and in the intrinsic case the inhomogeneity introduces dephasing during the storage operation that tends to lead to lower efficiencies upon retrieval when compared to the homogeneous case.

Regardless of efficiency, broadband atomic ensemble quantum memories tend to possess short memory lifetimes, mostly for technical (rather than fundamental) reasons. Narrowband atomic $\ket{s}\rightarrow\ket{g}$ transitions are desirable and necessary for achieving long memory lifetimes, but in practice in the broadband regime either inhomogeneous broadening (e.g., motional dephasing) or intrinsic material constraints (e.g., phonon lifetimes) tend to limit memory lifetime. That said, recent techniques have been developed to alleviate inhomogeneous broadening of the storage state (see Ref.~\cite{finkelstein2021continuous} and Sec.~\ref{hotatom_sec}), which may allow for simultaneous high-efficiency, broadband, and long-lived quantum memory operation. Even short-lived photonic quantum memories are useful, however, for such applications as quasi-on-demand single-photon generation and multiphoton quantum-state preparation \cite{nunn2013enhancing,finkelstein2018fast}, short-timescale synchronization, mode conversion \cite{fisher2016frequency,saglamyurek2014integrated}, and local quantum processing \cite{campbell2014configurable}, among others. Each quantum application employing quantum memory places different constraints on acceptable memory performance, and for some applications short memory lifetime is not prohibitive.

\subsection{Metrics}\label{metric_sec}

In what follows we give a brief description of the relevant metrics for quantum memory performance that have been developed in the literature:

\textbf{Efficiency} -- Quantum memory efficiency describes the integrated intensity ratio of photons sent in to the memory, and those retrieved from it:

\begin{equation}
    \label{eff_eq}\eta = \frac{\int_{-\infty}^{\infty}d\tau \abs{A_\text{out}(\tau)}^2}{\int_{-\infty}^{\infty}d\tau \abs{A_\text{in}(\tau)}^2} = \frac{\int_{-\infty}^{\infty}d\omega \abs{A_\text{out}(\omega)}^2}{\int_{-\infty}^{\infty}d\omega \abs{A_\text{in}(\omega)}^2},
\end{equation}

\noindent where $A_\text{in}(\tau)$ [$A_\text{in}(\omega)$] and $A_\text{out}(\tau)$ [$A_\text{out}(\omega)$] describe the incident and retrieved photon amplitude in the temporal (spectral) domain, respectively. 

Memory efficiency in atomic ensembles is determined by two processes, storage (or read-in) and retrieval (or read-out), which each have independent efficiencies $\eta_\text{stor}$ and $\eta_\text{ret}$, respectively, where $\eta = \eta_\text{stor} \eta_\text{ret}$. These efficiencies correspond to the integrated intensity ratio of photons sent in to the memory and the population entering the collective atomic storage state, $\abs{B_\text{out}(z)}^2$, and the integrated intensity ratio of photons retrieved from the memory and the same collective atomic state:

\begin{equation}
    \eta_\text{stor} = \frac{\int_{0}^{1}dz \abs{B_\text{out}(z)}^2}{\int_{-\infty}^{\infty}d\tau \abs{A_\text{in}(\tau)}^2}, \hspace{5em} \eta_\text{ret} = \frac{\int_{-\infty}^{\infty}d\tau \abs{A_\text{out}(\tau)}^2}{\int_{0}^{1}dz \abs{B_\text{out}(z)}^2},
\end{equation}

\noindent where $z=1$ in normalized units refers to the full length of the atomic ensemble, and where again spectral derivatives may be substituted for the temporal derivatives with impunity.

An important figure of comparison for memory efficiency is the optimal efficiency at a given optical depth, $d$. Described in Refs.~\cite{nunn2008quantum,gorshkov2007universal,gorshkov2007photon_2}, available atom number and optical depth of an atomic ensemble impose an upper limit on the achievable memory efficiency of a given system. The optimal bound on storage efficiency, $\eta_\text{opt}$, is calculated by finding the largest eigenvalue of the antinormally ordered storage kernel

\begin{equation}
    K(z,z') = \frac{d}{2} e^{-d(z+z')/2}I_0(d\sqrt{z z'}),
\end{equation}

\noindent where $I_0(x)$ is the zeroth order modified Bessel function of the first kind. The optimal bound on retrieval efficiency is also $\eta_\text{opt}$. Typically, optimal retrieval is only achievable for backward retrieval of the signal field \cite{gorshkov2007universal}; forward retrieval of the signal field typically suffers from reabsorption loss, as the signal field is reabsorbed by the atoms as it propagates through the ensemble after retrieval \cite{nunn2008quantum}. The optimal bound on total memory efficiency is $\eta_\text{opt}^2\geq\eta$.

\textbf{Bandwidth} -- Memory bandwidth corresponds to the full width at half maximum of the signal photon spectral intensity stored and retrieved in the memory. For a Gaussian-shaped signal field amplitude $A_\text{in}(\tau) \propto e^{-\tau^2/4\sigma^2}$, where $\sigma = \tFWHM / (2\sqrt{2 \ln 2})$ in terms of the signal temporal intensity full width at half maximum ($\tFWHM$), the Fourier-transform-limited photon bandwidth is $BW = 2\ln2/(\pi \tFWHM)$. The bandwidth of a quantum memory determines the memory's compatibility with short-duration pulses, which is of critical importance to real-world quantum applications that benefit from large clock rates and high processing speeds  \cite{reim2010towards,cacciapuoti2019quantum}.

\textbf{Memory Lifetime} -- The lifetime of an atomic ensemble quantum memory, $T$, typically corresponds to the time it takes for the retrieved photon population [$\langle A_\text{out}\rangle \propto \int_{-\infty}^{\infty}d\tau \abs{A_\text{out}(\tau)}^2$] to reach $1/e$ of its maximum value. Assuming the retrieval control field pulse remains unchanged aside from its arrival time in the atomic ensemble, this is equivalent to the time it takes for the the collective atomic state population [$\langle B_\text{out}\rangle \propto \int_{0}^{1}dz \abs{B_\text{out}(z)}^2$] to reach $1/e$ of its maximum value. Occasionally, the $1/2$ lifetime of a memory is given instead of the $1/e$ lifetime and a conversion factor depending on the decay model must be used to compare memories (e.g., for exponential decay, the $1/e$ lifetime $T$ and $1/2$ lifetime $T_{1/2}$ are related by $T_{1/2} = T \ln 2$; for Gaussian decay, $T_{1/2} = T \sqrt{\ln 2}$).

\textbf{Time-Bandwidth Product} -- A relevant figure of merit for most quantum protocols is the time-bandwidth product, defined as the product of memory lifetime and spectral bandwidth, $\mathrm{TBP} = T \times BW \pi/(2\ln 2)$. This figure provides a metric for the fractional delay produced by the memory in units of the photon duration, $\mathrm{TBP} = T / \tFWHM$. Broadband quantum memories may have equivalent or even larger time-bandwidth product at short storage times compared to narrowband quantum memories with long storage times.

\textbf{Noise} --  Noise is a critical figure of merit for determining the performance of a quantum memory. Even if the photon retrieved from a quantum memory is retrieved in the same quantum state as the input with unit efficiency, the presence of noise photons in addition to the retrieved signal renders such a memory useless for almost every quantum application. This importance of noise relates to whether a memory qualifies as `quantum'---true `quantum' memories must demonstrate below (ideally far below) 1 noise photon per retrieved signal photon, such that they do not contaminate the quantum states they store.

\begin{figure}[t]
	\centering
	\includegraphics[width=1\linewidth]{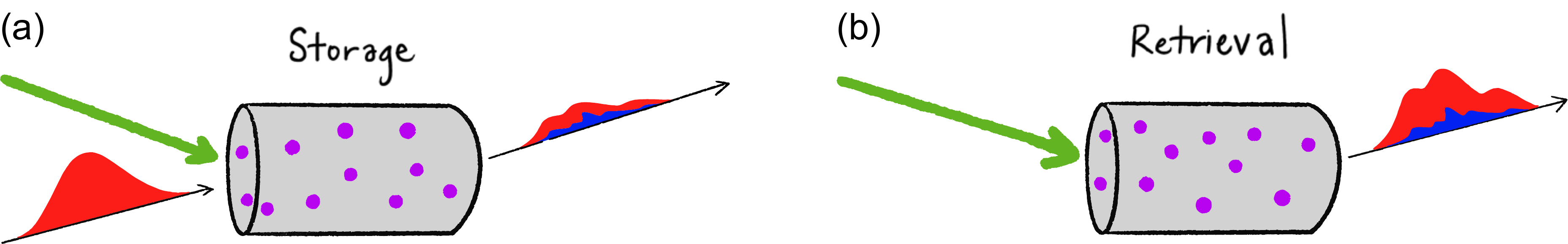}
	\caption{Schematic of atomic ensemble quantum memory during the (a) storage and (b) retrieval stages. The classical control field is shown in green, the quantum signal field in red, and the noise field in blue.}
	\label{fig_memoryschematic}
\end{figure}

The noise performance of quantum memory can be quantified several different ways. The most popular metrics for determining noise performance are: $\langle n_\text{noise}\rangle$, the average number of noise photons produced per pulse that overlap with the retrieved signal field; signal-to-noise ratio at 1 input photon/pulse (SNR), which is the signal-to-noise-ratio of the retrieved signal population to the average number of noise photons [$\mathrm{SNR} = \langle A_\text{out}\rangle / \langle n_\text{noise}\rangle = \eta\langle A_\text{in}\rangle / \langle n_\text{noise}\rangle = \eta / \langle n_\text{noise}\rangle$], evaluated when on average 1 photon per pulse is sent into the memory ($\langle A_\text{in}\rangle = 1$); $\mathcal{F}_\text{noise}$, the single-photon fidelity of the memory, defined as $\mathcal{F}_\text{noise} = 1 - 1/(\mathrm{SNR}+1)$, which is a measure of the fidelity of retrieving the stored signal photon rather than a noise photon, and which may be approximated as $\mathcal{F}_\text{noise} = 1 - 1/\mathrm{SNR}$ in the limit of large $\mathrm{SNR}$; and $\mu_1$, which is the ratio of the average number of noise photons per pulse to the memory efficiency, $\mu_1 = \langle n_\text{noise}\rangle / \eta = 1/\mathrm{SNR}$.

Some caution must be exercised in using each of these metrics, as ambiguity exists in the definition of $\mathrm{SNR}$. As shown in Fig.~\ref{fig_memoryschematic}(b), the total photon field retrieved from a memory is made up of two contributions, which we call the signal (shown in red) and the noise (blue). Occasionally, the \textit{total} photon field retrieved (signal and noise) is referred to as the ``signal,'' in which case the signal-to-noise ratio is reported as what we would consider the `total-to-noise' ratio ($\mathrm{TNR}$). The two are related by $\mathrm{TNR} = \mathrm{SNR}+1$.

\textbf{Fidelity} -- Independent from the noise performance of a memory, one can consider the overlap of the retrieved photon quantum state with the quantum state of the photon before it was sent into the memory. The definition of this fidelity typically depends on the degree of freedom of interest where quantum information is encoded, as in the case of quantum process tomography for polarization-encoded photonic qubits \cite{england2012high}, and can be performed for a single input state or averaged over multiple input states \cite{he2009dynamical}. Typically, fidelity is defined in terms of a state overlap $\mathcal{F} = \abs{\bra{\Psi_\text{out}}\Psi_\text{in}\rangle}^2$ for pure states or $\mathcal{F} = \bra{\Psi_\text{in}}\rho_\text{out}\ket{\Psi_\text{in}}$ for mixed states, which, in the temporal domain, can be expanded as

\begin{equation}
    \label{fidelityeq}\mathcal{F} = \frac{\abs{\int_{-\infty}^{\infty}d\tau \, A_\text{out}^*(\tau-\tau_d)A_\text{in}(\tau)}^2}{\int_{-\infty}^{\infty} d\tau \, \abs{A_\text{in}(\tau)}^2 \int_{-\infty}^{\infty} d\tau \, \abs{A_\text{out}(\tau)}^2},
\end{equation}

\noindent where $\tau_d$ is the time delay between incident and retrieved pulses. This definition of fidelity is sometimes called the ``waveform likeness'' \cite{chen2013coherent,wang2019efficient,wei2020broadband}, as it compares the likeness of the incident and outgoing temporal waveforms, where the outgoing waveform typically encounters some distortion [see Fig.~\ref{fig_memoryschematic}(a) and (b)]. In experiment, this is typically measured via intensity interferometry, where a reference pulse identical to the incident signal field is interfered with the retrieved pulse. Often in this case the interferometric visibility is reported in place of fidelity \cite{reim2010towards,saglamyurek2018coherent}.

Fidelity is a critical figure of merit as it relates to the degree of phase- and entanglement-preservation of a memory. Phase- and entanglement-preservation are critical to the use of quantum memories in quantum repeaters \cite{sangouard2011quantum}, for example, among other applications.

\textbf{Adiabaticity} -- The adiabaticity of a free-space atomic ensemble quantum memory is given by $d\tFWHM\gamma$, where $d$ and $\tFWHM$ are as defined above, and $\gamma$ is the coherence decay rate of the intermediate excited atomic state of the three-level atomic system ($\Gamma=2\gamma$ is the population decay rate). Described in detail in \cite{gorshkov2007photon_2,gorshkov2007universal,gorshkov2007photon_1}, quantum memories which satisfy $d\tFWHM\gamma\gg1$ are considered adiabatic, and memories with $d\tFWHM\gamma\sim1$ are typically considered non-adiabatic. The adiabatic condition $d\tFWHM\gamma\gg1$ ensures that for efficient memory operation the signal field bandwidth ($\propto 1/\tFWHM$) will be smaller than, or a similar order to, the excited state linewidth. It also ensures that the control field used to mediate the memory operation will be weak and long in duration relative to the timescale of evolution of the atomic states. All of these conditions together allow for adiabatic elimination of the atomic polarization in the equations of motion described in Sec.~\ref{ensemble_sec}.

\textbf{Multimode Capacity} -- A quantum memory capable of storing and retrieving multiple photonic qubits independently at overlapping times is considered multimode, as opposed to a single-mode memory, which must store and retrieve a given photon and reinitialize before it is prepared to store and retrieve another. Atomic ensembles may demonstrate multimode capacity in a number of different degrees of freedom, including temporal mode \cite{nunn2008multimode}, angular wavevector \cite{surmacz2008efficient}, three-dimensional spatial position \cite{pu2017experimental}, time bin \cite{afzelius2009multimode}, and frequency bin \cite{sinclair2014spectral}.

Multimode capacity is typically defined relative to a reference efficiency, $\eta_\text{ref}$. The multimode capacity $N$ of a quantum memory corresponds to the largest number of independent modes that can each be stored with at least $\eta_\text{ref}$ efficiency.

\textbf{Sensitivity} -- The sensitivity of atomic ensemble quantum memory is a metric that has been discussed only recently \cite{shinbrough2022variance,teja2021studying}. Memory sensitivity relates to a memory's performance in the presence of experimental noise, including fluctuations and drift in memory parameters such as optical depth and atomic transition linewidth as well as in the optical control field parameters used in the memory interaction. A memory that is less sensitive is more robust to experimental noise, and vice versa. 

Memory sensitivity may be quantified in a number of different ways, and memory sensitivity may in principle refer to the sensitivity of any of the metrics of this section in the presence of experimental noise. The typical case, however, concerns changes in memory efficiency in the presence of noise. When short-timescale fluctuations lead to changes in memory efficiency, the memory sensitivity is given in terms of the variance of memory efficiency in the presence of fluctuations $\zeta$ in parameters $\mathcal{X}$

\begin{equation}\label{Vfluc}
    V^\text{fluc}_\eta\left(\overline{\mathcal{X}}\right) = V_{\zeta}[\eta(\overline{\mathcal{X}}+\zeta)],
\end{equation}

\noindent where $\overline{\mathcal{X}}$ is the mean parameter value averaged over a large number of fluctuations, and where $V_x[y(x)] = \int dx\, y^2(x)P(x) - [\int dx\, y(x)P(x) ]^2$ is the unconditional variance of $y$ obtained when $x$ fluctuates with some probability distribution $P(x)$. The resulting standard deviation in memory efficiency can then be reported as $\sigma_\eta^\text{fluc}(\overline{\mathcal{X}}) = \sqrt{V^\text{fluc}_\eta(\overline{\mathcal{X}})}$. 

When long timescale drift in experimental conditions lead to changes in memory efficiency, and only one parameter of the system drifts, the memory sensitivity can be calculated via the one-at-a-time (OAT) variance

\begin{equation}\label{OAT_var}
    V_\eta^\text{OAT}(\overline{\mathcal{X}}) = V_{\mathcal{X}}[\eta(\mathcal{X})],
\end{equation}

\noindent where $\mathcal{X}$ varies uniformly over a finite range, $\mathcal{X}\in[\mathcal{X}^\text{min},\mathcal{X}^\text{max}]$ centered on $\overline{\mathcal{X}}$. Again, the standard deviation $\sigma_\eta^\text{OAT}(\overline{\mathcal{X}}) = \sqrt{V_\eta^\text{OAT}(\overline{\mathcal{X}})}$ may be used to quantify the change in memory efficiency due to changes in $\mathcal{X}$. In the case where multiple experimental parameters $\mathcal{X}=(x_1,...,x_N)$ drift simultaneously, a \textit{global} variance-based sensitivity analysis is required, wherein the most prevalent sensitivity measure is the first-order Sobol' variance \cite{sobol2001global,sobol2005global,sobol1993sensitivity}

\begin{equation}\label{Vi}
    V_i = V_{x_i}\{E[\eta(\mathcal{X})\vert x_i]\},
\end{equation}

\noindent where the inner expectation value, $E[\cdot]$, corresponds to the mean of $\eta(\mathcal{X})$ when $\mathcal{X}$ is varied over all possible values in a finite range at fixed $x_i$. The outer variance then measures the variance of this mean with respect to changes in $x_i$. The first-order Sobol' sensitivity index for parameter $x_i$ can then be calculated as

\begin{equation}\label{Si}
	S_i = V_i/V_\text{tot},
\end{equation}

\noindent where $V_\text{tot}$ is the total variance $V_\mathcal{X}[\eta(\mathcal{X})]$ observed over the range of interest.

\textbf{Telecom Compatibility} -- Quantum memories are often proposed for use before, after, or inside networks of optical fiber. As linear absorption in standard silica optical fiber reaches a minimum at 1310 and 1550 nm \cite{miya1979ultimate}, the telecom O- and C-band, respectively, a quantum memory compatible with these wavelengths is desirable. If 1310 or 1550 nm operation is not possible, other less-standard telecommunications bands exist in the range of 1260-1675 nm, and wavelength compatibility within this range is more desirable than outside this range. Telcom compatibility is therefore typically a binary metric, designating whether or not a quantum memory is capable of storage and retrieval of photons between 1260 and 1675 nm center wavelengths, where special attention is given to 1310 nm and 1550 nm operation.

\textbf{Latency} -- As with any technical device, latency describes the amount of time required by a quantum memory for initialization and preparation before storage and retrieval of photonic quantum states is possible. A useful figure of merit for this aspect of quantum memory operation is the memory's duty cycle, defined as the ratio of the amount of time the memory is capable of storage and retrieval to the total cycle time necessary for initialization and subsequent memory operation \cite{leung2022efficient}. In general, a memory with a larger duty cycle (ideally 100\%) is preferable to a memory with a smaller duty cycle.

\textbf{Size, Weight, and Power (SWaP)} -- Of practical importance to the use of quantum memories in real-world quantum applications is the size, weight, and power consumption of the memory. Typically these metrics are not reported exactly in the scientific literature, and may be hard to estimate, but general trends can be intuited (such as, for example, that an atomic vapor cell held at a higher temperature will have a higher power consumption than a cooler one, and memories requiring high control field optical powers will have higher power consumption than those with low powers). In general, a memory with a smaller SWaP is preferable to a memory with a larger SWaP.

\textbf{Device Lifetime} -- The lifetime of a quantum memory device is also of practical importance. In the limiting case, a quantum memory that is single-use --- only capable of storing and retrieving one photonic quantum state before the device needs to be replaced --- has virtually no practical application. All quantum memories in the literature exceed this limit; however, some platforms tend to demonstrate longer device lifetimes than others and again general trends can be intuited based on the required temperatures, pressures, and experimental components.

\section{Protocols and Hardware}\label{ProtocolHardware_Sec}

\subsection{Atomic Quantum Memory Protocols}\label{protocol_sec}

Many protocols for quantum memory operation in atomic ensembles exist based on distinct and quite disparate physical mechanisms. This section provides a brief review of the ensemble memory protocols found in the literature to date. We comment on the advantages of each memory in terms of the metrics described above in Sec.~\ref{metric_sec}, focusing in particular on the bandwidth limitations of each protocol.

\subsubsection*{Electromagnetically Induced Transparency (EIT)}

Electromagnetically induced transparency (EIT) in atomic ensembles is a phenomenon characterized by optical transparency at wavelengths resonant with the $\ket{g}\rightarrow\ket{e}$ atomic transition (see Fig.~\ref{fig_levelstruc}), which normally exhibits significant attenuation, and an associated reduction in group velocity of the incident signal field \cite{fleischhauer2005electromagnetically}. This transparency and slowed group velocity is due to the presence of a second, strong optical control field which couples the excited and metastable atomic states ($\ket{e}\leftrightarrow\ket{s}$), where the $\ket{e}\rightarrow\ket{s}$ transition is dipole allowed but the $\ket{s}\rightarrow\ket{g}$ tranistion is forbidden. This is a quite general phenomenon; for the purposes of this work we focus on EIT harnessed for quantum memory, where the `slow light' of general EIT is transformed into `stopped light' via adiabatic attenuation of the control field \cite{fleischhauer2002quantum}.

The pulse sequence for this protocol is as follows: A control field of duration longer than the signal field ($\tctrl>\tFWHM$) enters the medium ahead of the signal field in time ($\Deltau<0$). This opens a spectral transparency window at the signal frequency. This window is then slowly closed after the signal field enters the medium via attenuation of the control field. The signal field is thereby compressed and trapped in the medium in a superposition of the metastable and ground states of the atoms. This protocol is described in the narrowband regime in Refs.~\cite{fleischhauer2005electromagnetically,fleischhauer2000dark,fleischhauer2002quantum,phillips2001storage,lvovsky2009optical,gorshkov2007photon_2}, and in the broadband regime in Refs.~\cite{rastogi2019discerning,wei2020broadband}. 

Numerous experimental implementations of the EIT protocol for quantum memory exist (a review of most EIT quantum memory implementations can be found in Refs.~\cite{heshami2016quantum,lvovsky2009optical,ma2017optical,lei2022electromagnetically,novikova2012electromagnetically}). Efficiencies as high as 92\% have been demonstrated \cite{hsiao2018highly}, as well as---separately---memory lifetimes verging on 1 minute \cite{dudin2013light,heinze2013stopped}. As EIT memory requires adiabatic elimination of the excited state however, either large optical depths or narrow bandwidths (relative to the excited state linewidth) are required to fulfill the adiabaticity criterion $d\tFWHM\gamma\gg1$ \cite{rastogi2019discerning,shinbrough2021optimization}. To date this has limited operation of EIT-based quantum memory to bandwidths of 170 MHz \cite{wolters2017simple} or less. 

Noise operation in EIT systems tends to be limited by either four-wave-mixing (FWM) or control field noise. FWM noise occurs when the strong control field operates off-resonantly along the $\ket{g}\rightarrow\ket{e}$ transition, generating a spontaneous Stokes or idler photon along $\ket{e}\rightarrow\ket{s}$, before operating again along the $\ket{s}\rightarrow\ket{e}$ transition, generating an anti-Stokes or `spurious signal' photon along $\ket{e}\rightarrow\ket{g}$, which overlaps with the retrieved signal photon in all degrees of freedom. Control field noise occurs when there is a small frequency difference between signal and control field, and thus spectral isolation of the signal field is difficult to attain. Compared to ATS and SR memory protocols (discussed below) at the same bandwidth, EIT exhibits larger FWM noise due to its comparatively larger requirements on optical depth and control field Rabi frequency \cite{rastogi2022superradiance,saglamyurek2021storing}. FWM noise in EIT is very well characterized \cite{phillips2011light,lauk2013fidelity,geng2014electromagnetically} and many technical solutions to both control field and FWM noise exist \cite{ma2018noise,england2015storage,zhang2014suppression,nunn2017theory,dkabrowski2014hamiltonian,thomas2019raman}. 

In ladder systems, the EIT protocol is often referred to as fast ladder memory (FLAME), and may include a small detuning from resonance \cite{finkelstein2018fast,finkelstein2021continuous,davidson2022fast}.

\subsubsection*{Autler-Townes Splitting (ATS)}

The Autler-Townes Splitting (ATS) memory protocol is closely related to the EIT protocol \cite{rastogi2019discerning,shinbrough2021optimization}, but typically operates most efficiently at lower optical depths and broader photon bandwidths, placing it in the non-adiabatic class of quantum memories ($d\tFWHM\gamma\sim1$). The physical mechanism of ATS quantum storage relies on dynamic control of the Autler-Townes doublet created in the $\ket{g}\rightarrow\ket{e}$ absorption profile in the presence of a strong control field. By generating a dynamic Autler-Townes doublet that scans across the full bandwidth of the signal field (which is typically of the same order as the excited state linewidth, $\Gamma$) as it propagates through the atomic ensemble, uniform attenuation of the signal field in frequency can be achieved that ensures coherent population transfer to the spin wave state \cite{saglamyurek2018coherent}.

The pulse sequence used to generate the appropriate dynamical Autler-Townes splitting consists of a control field pulse of similar duration to the signal field ($\tctrl\sim\tFWHM$) that arrives at the atomic ensemble contemporaneously with the signal field ($\Deltau\sim0$) and possesses net control field pulse area (overlapping with the signal field) of $2\pi$.

As the bandwidth of the signal field increases, the effective optical depth of the $\ket{g}\rightarrow\ket{e}$ transition decreases due to increasing necessary Autler-Townes splitting, and as the bandwidth of the signal field decreases, population in the atomic polarization state experiences increased decoherence due to the increased duration of the protocol \cite{saglamyurek2018coherent}. Both of these effects lead to decreased memory efficiency, and therefore lead to an optimal photon bandwidth for a given ATS memory that is typically of order $BW\sim\Gamma$ \cite{saglamyurek2018coherent,rastogi2019discerning,shinbrough2021optimization}. This leads to favorable memory bandwidth compared to the EIT protocol in the same memory; however, to date ATS memories have been limited experimentally to bandwidths of 20 MHz \cite{rastogi2019discerning} and below.

\subsubsection*{Superradiance (SR) Mediated Memory}
Superradiance (SR) mediated memory is another on-resonant non-adiabatic protocol, similar to the ATS protocol, but with distinct characteristics. The protocol \cite{rastogi2022superradiance} relies on the effect of superradiance \cite{rehler1971superradiance}, which is characterized by cooperative spontaneous emission in atomic ensembles. The superradiant quantum memory protocol has three stages: absorption, writing, and retrieval. In the absorption stage, the signal photon is absorbed linearly along the $\ket{g}\rightarrow\ket{e}$ transition of the ensemble and thereby prepares the ensemble in a timed-Dicke state \cite{roof2016observation} with a decay time shorter than the bare atomic excited state lifetime, as the induced ordered spatial phase distribution encourages the coherent enhancement of radiation in the direction of the absorbed photon. This short decay time makes SR memory compatible with photons with bandwidth greater than the bare excited state linewidth, $BW>\Gamma$. To suppress superradiant emission of the absorbed photon and realize photon storage, the writing stage starts directly after the absorption stage, within the superradiant decay time $T_{SR}$. This process is accomplished by sending in a control field with $\pi$ pulse area and duration $\tctrl \ll T_{SR}$, which maps the generated atomic polarization to a collective spin excitation. During the retrieval process, another $\pi$-pulse control field is applied and the photon is superradiantly emitted.

Optimizing the absorption process in SR memory requires shaping the temporal profile of the signal photon to match the time-reversed superradiant decay. The optimal shape of the photon is therefore exponentially rising with time constant inversely proportional to the linewidth $\Gamma$ and the optical depth $d$. Currently, the first demonstration of SR memory has achieved storage of signal photons with time constants down to $10 \text{ns}$ and $3\%$ memory efficiency \cite{rastogi2022superradiance}. The lower-than-expected efficiency is due to a long experimental control pulse duration $\tctrl$ that does not completely satistfy the condition $\tctrl \ll T_{SR}$. When compared with EIT and ATS protocols, the SR protocol requires less optical depth for the same memory efficiency at the cost of higher control power. Four-wave mixing noise in SR memory appears to be on par with EIT but much higher than ATS memory.

\subsubsection*{Absorb-then-Transfer (ATT)}

Similar to SR memory, the absorb-then-transfer (ATT) memory protocol \cite{moiseev2001complete,vivoli2013high,gorshkov2007photon_2,carvalho2020enhanced} occurs in three physically distinct stages. As the name suggests, the first two stages correspond to linear absorption along the $\ket{g}\rightarrow\ket{e}$ transition in the absence of the control field, then a short delay later the application of a $\pi$-pulse control field along the $\ket{e}\rightarrow\ket{s}$ transition that transfers population from the atomic polarization coherence ($\ket{g}\leftrightarrow\ket{e}$) to the spin wave coherence ($\ket{g}\leftrightarrow\ket{s}$). These two stages implement the storage operation. When retrieval is desired, another control field pulse with $\pi$ pulse area is applied to the atomic ensemble, transferring population back to the atomic polarization coherence, which then emits the stored signal field through dipolar radiation. This memory protocol is distinct from the photon-echo protocols discussed below as it relies on homogeneous broadening of the intermediate excited state rather than (reversible or structured) inhomogeneous broadening.

In order to optimize storage efficiency in the ATT protocol, the storage control field arrival should be synchronized in time with the first zero of the complex signal field amplitude evaluated at the middle of the ensemble ($L/2$) \cite{carvalho2020enhanced}. This protocol can implement optimal photon storage (i.e., $\eta_\text{stor}=\eta_\text{opt}$) for large optical depths, when the adiabaticity criterion $d\tFWHM\gamma\gg1$ is satisfied \cite{vivoli2013high}. As this protocol is typically employed when $\tFWHM\gamma<1$, this implies $d\gg1/(\tFWHM\gamma)$.

Near-off-resonant memory (NORM) operation of ATT was introduced in Ref.~\cite{shinbrough2022broadband} as a means to achieve higher efficiency than resonant ATT when control field power is constrained. NORM balances reabsorption loss, which is worst on resonance at large optical depths, and finite available control field power, which leads to lower efficiency at larger detuning. NORM operation is not necessary in the ideal case, for example, investigated in Ref.~\cite{shinbrough2021optimization}, when all parameters of the control field are optimized to improve memory efficiency. NORM operation only appears to help in the case when one parameter of the control field is constrained, as is frequently the case experimentally. Future work is needed to investigate this behavior theoretically.

The NORM ATT protocol is well-suited for broadband operation, as was recently demonstrated in Refs.~\cite{shinbrough2022broadband} and \cite{shinbrough2022efficient} in atomic barium vapor. A storage efficiency of 95.6$\pm$0.3\% was reached for ultrashort photons (500 fs), which, as noted in Sec.~\ref{bwlw_sec}, is the highest attained for bandwidths $>$10 MHz (see Fig.~\ref{fig_eff_v_bw} in Sec.~\ref{eff_sec}). The total efficiency was measured to be 31$\pm$1\% at 900~$^{\circ}$C, limited by control field power. The lifetime of this memory was measured to be 0.49(1) ns at 900~$^{\circ}$C, which, while not long in absolute terms, is significantly better than the $\sim$ps level lifetimes that are typical of solid-state THz memories (see Sec.~\ref{lifetime_sec}), which are usually assumed to be population-lifetime limited; the bare atomic 0.25 sec population lifetime of the storage state used in atomic barium leaves large room for improvement. This lifetime corresponds to a time-bandwidth product of $\mathrm{TBP} = 980\pm20$. In terms of noise performance, a signal-to-noise ratio of $\mathrm{SNR}=$(8.2$\pm$1.3)$\times10^3$ is measured for an input of 1 photon per pulse, leading to a single photon fidelity of $F = 0.99988(2)$. This noise performance is on par with the lowest noise ladder-type atomic ensemble memory systems \cite{kaczmarek2018high,finkelstein2018fast,thomas2022single,davidson2022fast}, as shown in Sec.~\ref{noise_sec}. The exceptional noise performance of this memory is explained by the ultra-large ground-storage state splitting, which is so large ($\sim$340 THz) that it eclipses the excited-storage state splitting ($\sim$200 THz), the transition operated along by the control field. This means that, to first order, the control field frequency is insufficient to excite four-wave mixing noise, which is the dominant noise contribution in almost all other $\Lambda$-type memories. This large ground-storage state splitting also virtually eliminates noise from thermal population of the storage state, and furthermore allows for near complete spectral suppression of control field leakage noise into the signal path.

\subsubsection*{Off-Resonant Raman}
The off-resonant Raman memory protocol \cite{nunn2007mapping} is well-suited for storage and retrieval of broadband photons. In a typical $\Lambda$-type or ladder system, the strong control field, which is off-resonant from the excited state $\ket{e}$ by a detuning $\Delta$ (see Fig.~\ref{fig_levelstruc}), enables a two-photon transition between the ground state $\ket{g}$ and storage state $\ket{s}$. In the storage process, the signal photon, which is also detuned by $\Delta$ from the $\ket{g}\rightarrow\ket{e}$ transition, is spatially and temporally overlapped with the control field and is thereby mapped onto a collective spin-wave excitation. A subsequent retrieval field (typically the same as the storage field) is then applied to the atomic ensemble and the stored photon is deterministically released, completing the retrieval process. 

In the off-resonant Raman limit, the detuning is much greater than the excited state linewidth, the bandwidths of signal and control fields, and the control field Rabi frequency $\abs{\Delta}\gg \Gamma,BW,\abs{\Omega(\tau)}$. This allows the off-resonant Raman protocol to operate in the adiabatic regime where the excited state is eliminated. In principle, this may also allow for the storage and retrieval of photons with arbitrary temporal shapes, and for arbitrary temporal shaping upon retrieval, depending on the shape of the control field. The memory efficiency for this protocol is proportional to the optical depth and control pulse energy and is inversely proportional to the detuning. The bandwidth of storage is typically only limited by the energy splitting between ground and storage states, and has therefore exceeded 1 THz in experiment \cite{england2015storage,fisher2017Storage,england2013photons,bustard2013toward}. Due to the large detuning, four-wave mixing noise is typically dominant over control field leakage noise and other noise contributions. 

In ladder systems, the off-resonant Raman protocol with few modifications is referred to as off-resonant cascaded absorption (ORCA) \cite{kaczmarek2018high,thomas2022single}.

\subsubsection*{Controlled Reversible Inhomogeneous Broadening (CRIB)}

In atomic ensembles that exhibit inhomogeneous broadening of the optical transition, the atomic coherence generated by the absorption of the incoming signal will begin to rapidly dephase. Controlled reversal of the inhomogeneous broadening at a time $t$ following absorption leads to time-reversal of this dephasing and the subsequent re-emission of the signal field at time 2$t$ \cite{campbell2019echo, sangouard2007analysis, kraus2006quantum, moiseev2001complete,lvovsky2009optical,simon2010Quantum}. A review of the equations of motion governing this process can be found in Sec.~\ref{MB_sec} of this chapter and in the proposal of this protocol, Ref. \cite{sangouard2007analysis}. This memory protocol is commonly referred to as controlled reversible inhomogeneous broadening (CRIB). The bandwidth for this protocol is set by the inhomogeneous broadening of $\ket{g}\rightarrow\ket{e}$ transition \cite{campbell2019echo, mcauslan2011Photonecho}.  A subset of these protocols are the gradient echo memories (GEM) in which the controlled inhomogeneous broadening is given by a spatial electric or magnetic field gradient \cite{campbell2019echo}. 

The efficiency of CRIB is derived in Refs.~\cite{sangouard2007analysis,simon2010Quantum}. For a narrow absorption feature of initial linewidth $\gamma_{0}$ that has been inhomogeneously broadened to $\gamma$, the efficiency of CRIB is given by: 

\begin{equation} 
    \eta_\text{CRIB} = (1 - e^{-d\gamma_{0}/\gamma})^2
\end{equation}

\noindent where $d$ is the optical depth of the unbroadened line. The efficiency goes as the square of the probability of absorption, as the retrieval process is given by time-reversed absorption \cite{simon2010Quantum, moiseev2007photon}. Additional considerations and expressions for the efficiency are discussed in Refs.~\cite{lvovsky2009optical,longdell2008Analytic}. To date, the bandwidth and efficiency of experimental implementations of the CRIB protocol have been limited to $\sim0.7$ MHz and 69\%, respectively \cite{hedges2010efficient}.

\subsubsection*{Gradient Echo Memory (GEM)}

Gradient echo memories (GEM) are an implementation of the CRIB protocol in which the inhomogeneous broadening is given by a (typically spatial) electric or magnetic field gradient \cite{campbell2019echo,lvovsky2009optical,simon2010Quantum}. In these schemes, the spatial gradient gives rise to a spatially dependent detuning that can be reversed by changing the sign of the gradient. The GEM protocol consequently does not require backward propagation of the signal for high efficiency \cite{campbell2019echo}. GEM was first implemented at the few-photon level over a decade ago using light in the telecommunications C-band, at a bandwidth of 5 MHz, with a storage time of several hundred nanoseconds \cite{lauritzen2010telecommunication}. Implementations in cold atomic ensembles have also been demonstrated over the last decade \cite{Leung2022observation, Sparkes2013an, sparkes2010ac}, however the bandwidths demonstrated (sub-MHz) tend to be small due to the difficulty in generating GHz-THz inhomogeneous linewidths.

\subsubsection*{Atomic Frequency Comb (AFC)}

\begin{figure}[t]
    \centering
    \includegraphics[scale=0.2]{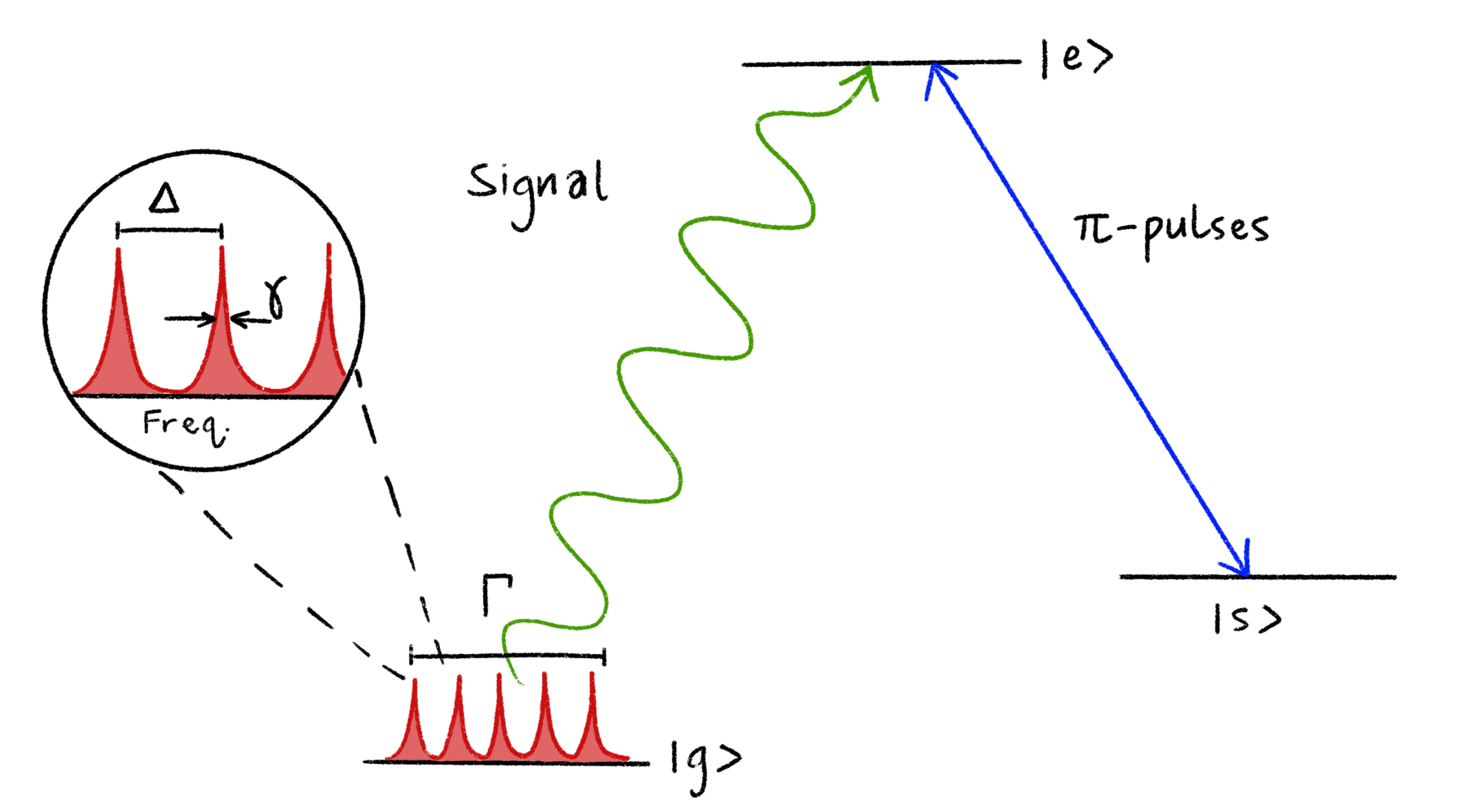}
    \caption{Schematic of the atomic frequency comb (AFC) memory. A periodic comb structure of total width $\Gamma$ is formed in the inhomogeneous line. The comb teeth have linewidth $\gamma$ and are evenly spaced by $\Delta$. More details can be found in Ref.~\cite{afzelius2009multimode}.
    }
    \label{fig_AFC}
\end{figure}

The atomic frequency comb (AFC) protocol is an inherently multimode photon echo memory that utilizes many narrow, periodically spaced absorption features created in a broad inhomogeneous $\ket{g} \rightarrow \ket{e}$ line by optical pumping \cite{de2008solid, afzelius2009multimode, lvovsky2009optical, campbell2019echo, heshami2016quantum,simon2010Quantum}. Contrary to the CRIB and GEM protocols, the periodic structure of this frequency comb itself leads to periodic rephasing, without the need for external fields to reverse the inhomogeneity. The periodic structure has total width $\Gamma$, and is composed of narrow comb teeth of linewidth $\gamma$ evenly spaced in frequency by $\Delta$. The finesse of the comb teeth is then $\mathscr{F} = \Delta/\gamma$. Since atoms are selectively removed from the inhomogeneous line to form the comb structure, the optical depth of the transition has been reduced approximately by a factor of $\mathscr{F}$ \cite{de2008solid, afzelius2009multimode}.

Signal bandwidths compatible with the AFC protocol must satisfy $\Delta < BW \lesssim \Gamma$. Upon absorption of the signal field, the periodic structure of the frequency comb causes the atomic coherence to rephase after a time $2\pi/\Delta$, leading to subsequent reemission of the signal field. The AFC protocol thus possesses a fixed, predetermined storage time set by the spacing of the comb teeth; however, prolonged and arbitrary storage times can be achieved by mapping the coherence to a third metastable state, $\ket{s}$, via optical $\pi$-pulses \cite{afzelius2009multimode,lvovsky2009optical, campbell2019echo}, as shown in Fig.~\ref{fig_AFC}. 

Retrieval of the signal field in the forward direction is limited to a maximum efficiency of 54\% due to reabsorption of the signal field by the medium \cite{afzelius2009multimode, lvovsky2009optical, campbell2019echo}, and is given by the closed form $\eta_\text{AFC} \approx (d/\mathscr{F})^2 e^{-7/\mathscr{F}^2}e^{-d/\mathscr{F}}$. Retrieval of the signal field in the backward direction is not subject to this constraint, and is given by $\eta_\text{AFC} \approx (1 - e^{-d/\mathscr{F}})^{2}e^{-7/\mathscr{F}^{2}}$ \cite{afzelius2009multimode}.

A distinct advantage of the AFC protocol is that the total number of temporal modes that can be stored does not depend on the optical depth of the medium, but rather the total number of comb teeth that are prepared \cite{simon2010Quantum}. In contrast, the capacity of the EIT protocol to store $N$ modes scales as $\sqrt{d}$, and for the CRIB protocol, for sufficiently large optical depth, the number of modes $N$ scales linearly with $d$ \cite{afzelius2009multimode, simon2010Quantum, lvovsky2009optical}. 

Out of those protocols employing an inhomogeneously broadened $\ket{g}\rightarrow\ket{e}$ transition, the AFC protocol tends to be the most broadband, where bandwidths up to 50 GHz at an efficiency of 1.8\% have been demonstrated in the literature \cite{saglamyurek2016multiplexed}.

\subsubsection*{Revival of Silenced Echo (ROSE)}

\begin{figure}[t]
    \centering
    \includegraphics[scale = 0.2]{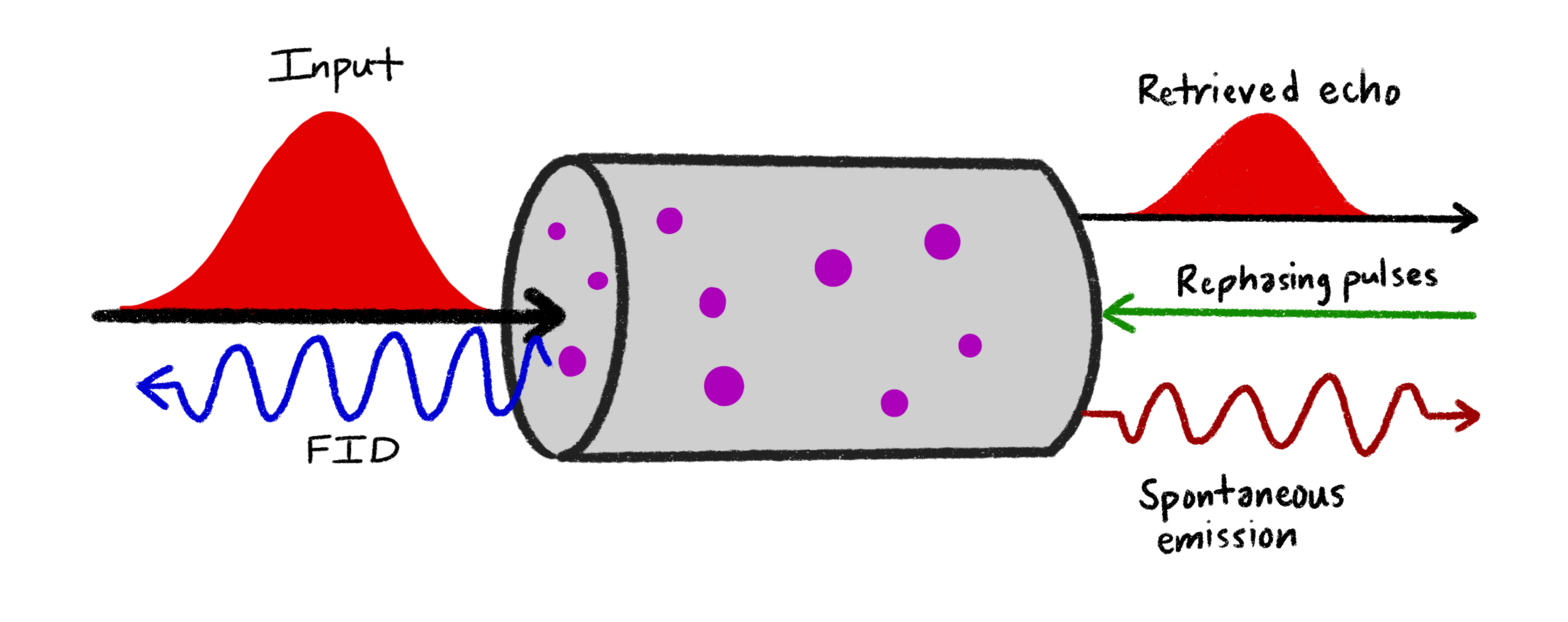}
    \caption{Schematic of the ROSE protocol outlined in Ref.~\cite{bonarota2012revival}. The rephasing pulses (green) are counterpropagating relative to the input signal. In this geometry, the free induction decay (blue) will be emitted in the same direction as the rephasing beams. Spontaneous emission noise (thin red) will still be present and pollute the retrieved echo emitted in the forward direction.}
    \label{fig:rose}
\end{figure}

The two-pulse photon echo is a natural candidate for quantum memory in inhomogeneously broadened media \cite{ruggiero2009two}. In principle, two-pulse photon echoes do not require any state preparation, and may utilize the entire inhomogenous linewidth and optical depth, which may improve the multimode capacity of such a memory compared to the AFC protocol \cite{dajczgewand2014large, ruggiero2009two,damon2011revival}. Such a protocol can in principle also be extended to a three-level system to transfer the inhomogeneous atomic coherence to a longer-lived spin state \cite{damon2011revival}. A two-photon echo on its own is not suitable for quantum memory, however, as population inversion from the strong rephasing pulse creates significant and unavoidable spontaneous emission and free induction decay (FID) noise \cite{ruggiero2009two,damon2011revival}. 

A modification to the two-pulse photon echo protocol, the revival of silenced echo (ROSE) protocol, is able to suppress some of these inherent noise sources and provide quantum operation. This protocol relies on phase-mismatching of the incoming signal and the first rephasing pulse. The atomic polarization in the medium persists regardless of the phasematching between these two fields, and phasematching with a second rephasing pulse enables coherent emission of a second photon in the absence of FID noise. The total storage time for this process is $2(t_{2} - t_{1})$, where $t_{1}$ and $t_{2}$ are the arrival times of the first and second rephasing pulses, respectively \cite{damon2011revival, bonarota2012revival,bonarota2014photon}. The geometry of this scheme is described in Refs.~\cite{damon2011revival, bonarota2012revival,bonarota2014photon} and illustrated in Fig. \ref{fig:rose}. If both rephasing pulses are co-linear and counterpropagating relative to the incoming signal, FID will be emitted in the direction of the rephasing pulses and not the retrieved echo. Spontaneous emission noise cannot be eliminated and remains a limitation to the noise performance of this protocol. 

In the limit of an infinite $T_2$ coherence time, and assuming forward emission of the second echo, the efficiency of ROSE depends solely on the optical depth as $\eta_\text{ROSE} = d^2 e^{-d}$. For a finite $T_{2}$, the efficiency also depends on $T_{2}$ and the time difference between the two rephasing pulses as $\eta_\text{ROSE} = d^{2} e^{-d} e^{-4(t_{2}-t_{1})/T_2}$ \cite{damon2011revival, bonarota2012revival}.

In the proposal for the ROSE scheme, the authors note that $\pi$-pulses are not optimal for the population inversion \cite{damon2011revival}; adiabatic passage with chirped pulses is preferable and has been used in several experimental implementations of ROSE \cite{dajczgewand2014large, damon2011revival,  bonarota2012revival, minnegaliev2022implementation}. To date, the ROSE protocol has been exclusively implemented in rare-earth ion-doped solids at cryogenic temperatures: Tm$^{3+}$:YAG \cite{bonarota2014photon,bonarota2012revival,minnegaliev2018experimental,gerasimov2017quantum}, Er$^{3+}$:YSO, \cite{dajczgewand2014large, minnegaliev2022implementation}, and Tm$^{3+}$:YSO \cite{minnegaliev2021linear}. Efficiencies as high as 44\% have demonstrated in 0.005\% $^{167}$Er$^{3+}$:YSO, as well as storage times of up to 16 $\mu$s, though memory bandwidths are typically limited to below 1 MHz \cite{minnegaliev2022implementation,dajczgewand2014large}.

\subsubsection*{Hybrid Photon-Echo Rephasing (HYPER)}

Hybrid photon-echo rephasing (HYPER) is a protocol that combines the concepts of CRIB with a two-photon echo, thus avoiding the optical pumping step required for protocols like CRIB and AFC, and reducing the noise associated with the na{\"i}ve two-pulse photon echo protocol. In HYPER, an inhomogeneously broadened atomic ensemble is further inhomogeneously broadened in a controllable fashion after absorption of the signal field. The controllable broadening is then turned off, a $\pi$-pulse typical of the two-pulse photon echo scheme is applied, and the controllable broadening is turned back on. By introducing this controllable broadening, the standard two-pulse echo is silenced, and after turning off the controllable broadening a second time, and applying a second $\pi$-pulse, finally the atomic polarization rephases and the signal field is reemitted. This scheme ultimately reduces the number of atoms that would experience gain and undergo spontaneous emission, thus reducing the noise inherent to two-photon echoes \cite{mcauslan2011Photonecho}. Analytical treatment of the protocol, and the first experimental implementation with 0.2 MHz bandwidth and $\sim$10\% efficiency, can be found in Ref.~\cite{mcauslan2011Photonecho}.

\subsection{Hardware}

We now move on to discuss the various types of hardware used for atomic ensemble quantum memory. Each type possesses unique advantages and disadvantages, and may place limits on the memory protocols that are possible to implement in each system (e.g., rare-earth ions doped in solids with intrinsic inhomogeneous broadening).

\subsubsection{Hot Atoms}\label{hotatom_sec}

Hot atomic ensembles have been widely used to establish different quantum memory schemes, thanks to their advantages of room (or near-room) temperature operation and reduced experimental complexity. Due to their relatively simple electronic structure, high vapor pressure, and long-lived spin states, the alkali metals rubidium and cesium have been the atomic species most widely used for hot atomic ensemble quantum memory. The recently demonstrated barium memory is an exception, where instead of spin states, electronic orbitals were used for the memory interaction to enable THz-bandwidth storage \cite{shinbrough2022broadband, shinbrough2022efficient}. In addition to hot atomic species, molecular species in room-temperature gas cells have also been used for ensemble quantum memory \cite{bustard2013toward,bustard2017quantum}, where a phononic storage state is used instead of a hyperfine spin level as is the case for the alkali atoms.

Glass vapor cells have the advantage of widespread commercial availability, and are thus one of the most frequently used hardware implementations for hot atomic ensemble quantum memory. These cells are prepared using precision scientific glassblowing techniques, in which the atoms are introduced to the glass cell under vacuum conditions. One disadvantage of glass vapor cells are the inelastic collisions between atoms inside the cell and the cell walls, which can limit memory lifetimes in the narrowband regime \cite{jiang2009observation} or when the atoms are optically pumped into the metastable spin state \cite{guo2019high}. These collisions occur on the microsecond timescale for a typical, centimeter-size cell, and can further induce spin depolarization, thus reducing the lifetime of any collective spin state, including those used in ensemble quantum memory. Buffer gases such as Ar, Ne or N$_2$ are commonly loaded into the vapor cell along with the atomic species of choice to increase the lifetime of collective spin states and to reduce the broadening of certain transitions due to inelastic collisions \cite{brandt1997buffer}. As it slows atomic diffusion and limits the absorption of atoms onto the cell walls, the presence of a buffer gas may also increase the device lifetime of glass vapor cells. Unfortunately, the presence of buffer gas can also induce additional spin depolarization and quenching due to collisions between the atoms and buffer gas \cite{manz2007collisional}. An optimal buffer gas pressure can be found that balances mitigation of wall collisions and the increase in atom--buffer-gas collisions \cite{happer1972optical}. Paraffin-coated vapor cells have also been studied as an alternative method to increase collective spin state lifetimes by inducing elastic instead of inelastic wall collisions \cite{bouchiat1966relaxation,jiang2009observation}. A disadvantage of paraffin-coated or similar vapor cells is their complex fabrication process, as well as the fact that the coating material may be incompatible with elevated temperatures. Collisions between atoms and buffer gas can also be a source of noise due to collision-induced fluorescence \cite{manz2007collisional,rousseau1975resonance}. This noise source limits the fidelity of quantum memory, particularly at or near resonance, and is greatly reduced when the quantum memory is operated off resonance and optical filtering is applied. Buffer-gas free vapor cells naturally suppress collision-induced fluorescence noise \cite{jiang2009observation}. 

The lifetime of collective spin states is not only influenced by collisions. Broadband quantum memories using hot atomic ensembles may also suffer from spin-wave dephasing arising from thermal atomic motion \cite{zhao2009millisecond}. This atomic motion may result in atoms wandering out of the effective interaction region defined by the laser fields, or it may result in scrambling of the spin-wave phase as the constituent atoms move within the interaction region. The rate of this latter dephasing process is proportional to $\Delta \vec{k} \cdot \vec{v}$ with $\Delta \vec{k} = \vec{k}_s - \vec{k}_c$ representing the wavevector of the spin wave, where $\vec{k}_s$ and $\vec{k}_c$ are the wavevectors of signal and control fields, respectively, and $\vec{v}$ representing the thermal velocity of the atoms. In general, it is desirable generate a spin wave with long wavelength in order to maintain spin-wave coherence. This can be modified somewhat through experimental design \cite{jiang2009observation}, but in many cases the spin-wave wavelength is limited by the large detuning of the signal and control fields. One scheme \cite{finkelstein2021continuous} has demonstrated elimination of motional dephasing through dressing of the collective spin state to an auxiliary `sensor' state. A sensor state corresponds to a state that, when coupled via an optical field to the storage state, exhibits opposing energy shifts under the same broadening mechanism. First demonstrated experimentally in a Doppler-broadened system \cite{finkelstein2021continuous}, off-resonant fields couple the storage state to the sensor state, dressing the collective excitation such that the frequency shifts due to the dephasing mechanism cancel out. Thus, the dephasing of the spin wave is protected at the price of one or two additional dressing fields. Another scheme using velocity-selective optical pumping \cite{main2021preparing} has been proposed and proved to be effective against motional dephasing at the cost of reduced optical depth. In this scheme the atoms are pumped out from ground state to an auxiliary state and then pumped back to the ground state using a narrowband laser, which selectively narrows the motional distribution of the atoms. The rate of dephasing is therefore reduced owing to the small $\Delta\vec{v}$ of the atoms. Other mechanisms that induce spin wave dephasing include spatial and temporal variation of longitudinal magnetic field. Note that this is not unique to hot atomic ensembles and experiments have been performed to show prolonged spin-wave coherence using hot atomic vapors through magnetic shielding \cite{appel2008quantum} or applying a strong guiding magnetic field \cite{specht2011single,tian2015suppressing}.

The heat-pipe oven is an alternative to glass vapor cells, and represents a more robust, higher-temperature, customizable vapor cell \cite{vidal1969heat,vidal1971heat,vidal1996vapor}. Typically heat-pipe ovens are comprised of a stainless-steel tube covered by a heater with water cooling at both ends. Inside the tube, the wall is covered by mesh that acts as wick. Upon heating, atoms from liquid metal evaporate, form a dense vapor, and diffuse toward both ends of the oven until they condense in the water-cooled region and return to the center through the wick due to the capillary effect. The heat-pipe oven can be used to establish a homogeneous temperature and density distribution oven for a long period of time. It can handle high temperatures, which is suitable for low-vapor-pressure atomic species or high-vapor-pressure species at much higher optical depth. The design of the oven can prevent vapor from condensing on the cell windows, and thus the windows tend to become opaque more slowly than in equivalent glass vapor cells. The species of atoms and buffer gas can also be changed more readily than in sealed glass vapor cells.

Many efforts have been made to miniaturize the hot atomic vapor cell in order to achieve chip-scale operation, enabling compact design and lower necessary optical power \cite{kitching2018chip}. Early attempts to integrate hot atoms with optical fibers and silicon platforms include loading hot atomic vapor near tapered fibers \cite{hendrickson2010observation}, into hollow-core fibers \cite{ghosh2006low}, or near waveguides \cite{yang2007atomic}. These approaches in principle allow long interaction length while maintaining tight confinement, although to-date only quantum memory in hollow-core fibers has been implemented experimentally \cite{sprague2014broadband}. The large core size of kagome fiber (26 $\mu$m) allows for a large optical depth compared to other approaches \cite{hendrickson2010observation}. The achieved memory lifetime of 100 ns could be further extended using buffer gas or a spin-preserving coating. Light-induced atomic desorption (LIAD) \cite{alexandrov2002light,karaulanov2009controlling} has been found effective in increasing optical depth without heating in miniaturized vapor cells and in kagome fiber \cite{talker2021light, ghosh2006low, sprague2014broadband}.

\subsubsection{Cold Atoms}

Similar to the case in hot atoms, the alkali metals are the species most commonly used for cold atomic ensemble quantum memory. Storage times in hot atomic vapors are limited by various mechanisms, including Doppler broadening and transit-time broadening, which are effectively eliminated in laser-cooled atomic ensembles. However, laser cooling tends to require large ultra-high vacuum chambers, multiple stabilized lasers for cooling and trapping, and typically imposes large latency and a non-unity duty cycle in order to re-cool atoms between shots. The colder the atoms are, the denser the cloud and the further the reduction in motion-related broadening and dephasing. However, the maximum length of the ensemble is typically reduced for colder temperatures due to practicalities of the various laser fields used for cooling and trapping and the process of evaporative cooling. As a result, there continue to be quantum memory demonstrations in cold atomic gasses in magneto-optical traps at $\sim\mu\rm{K}$ temperatures and in Bose-Einstein condensates at $\sim\rm{nK}$ temperatures. Optical depths as high as $\sim1000$ can be achieved \cite{hsiao2018highly}. We note that for very long-lived storage, an optical lattice along the k-vector of the stored spin wave is required to eliminate residual motional effects, which substantially reduces the available optical depth and largely precludes broadband storage.

\subsubsection{Solids}\label{solids_sec}

\subsubsection*{Optical phonon modes in diamond}
Over the last 10 years the use of optical phonon modes in bulk diamond has emerged as a platform for fast, broadband quantum memory. Such memories are attractive due to their THz bandwidth operation across the visible and near infrared due to the large bandgap of diamond, and room temperature operation \cite{england2013photons,kalish2007Diamond,england2015storage,england2016PhononMediated,fisher2017Storage,lee2012Macroscopic}. A $\Lambda$-system consisting of the ground state of diamond, an optical phonon mode storage state, and the conduction band as the excited state provides a storage bandwidth of up to 40 THz, limited only by the splitting between the ground state and the optical phonon mode. The memory lifetime in this platform is ultimately limited to 3.5 ps by the decay of the phonon mode \cite{england2013photons, lee2012Macroscopic}. 

The first proof-of-principle experiment to demonstrate the suitability of diamond as a quantum memory yielded a noise floor well below 1 photon per pulse \cite{england2013photons}. The primary noise source was FWM noise intrinsic to most $\Lambda$-type memory schemes; however, the authors highlight that FWM is suppressed due to the large dispersion in diamond \cite{england2013photons}. Other sources of noise, such as resonant fluorescence, are greatly suppressed by operating with a detuning of $\sim$950 THz from the conduction band \cite{england2013photons, england2015storage, fisher2017Storage}.

True single-photon storage was first demonstrated with THz-bandwidth (260 fs) photons produced from a heralded spontaneous parametric down conversion (SPDC) source \cite{england2015storage}. The preservation of nonclassicality in the retrieved photons was determined by measuring the second-order correlation function using Hanbury-Brown Twiss interferometry \cite{england2013photons, england2015storage}. Storage of polarization qubits and of a single photon from a polarization-entangled pair has been demonstrated with a fidelity of 76\% for picosecond-long storage \cite{fisher2017Storage}.

\subsubsection*{Rare-earth ions}
Trivalent rare-earth ions doped in solids at cryogenic temperatures have many properties that make them a promising platform for quantum memory devices, including long-lived coherence, compatibility with integrated photonics, lack of motional dephasing, high density of ions, moderate optical depths, and telecommunications wavelength compatibility \cite{ohlsson2002quantum,wesenberg2007scalable,hizhnyakov2021rare,roos2004quantum,walther2015high,ahlefeldt2020quantum,grimm2021universal,thiel2011rare,zhong2017interfacing}.  Optical coherence times can exceed milliseconds \cite{arcangeli2015temperature,bottger2009Effects} and, to date, the longest spin coherence time observed is 6 hours \cite{zhong2015optically}. 

The long optical and spin coherence properties afforded to the rare earths is due to their unique electronic structure. The optical transitions of interest are between their 4f-4f valence electronic levels, which are dipole-forbidden in the free ion. The 4f valence shell is electronically shielded by filled 5s and 5p orbitals, which renders the 4f levels largely insensitive to their environment. Incorporation of the rare-earth ion in a solid introduces a crystal electric field which acts as a perturbation to the free-ion states. For sufficiently low site-symmetry of the rare-earth ion, the crystal electric field mixes in additional states, which weakly allows electronic transitions \cite{liu2006spectroscopic,thiel2011rare}. 

The electronic states of the free ion are labeled by Russell-Saunders term symbols where the electronic spin $J$ and projection $m_{j}$ are considered good quantum numbers \cite{thiel2011rare, liu2006spectroscopic,ahlefeldt2013Evaluation}. These states are $(2J+1)$-degenerate, known as Kramer’s degeneracy. The addition of the crystal electric field either partially or completely lifts the degeneracy of these states. For sufficiently low site symmetry and ions with an even number of electrons---the non-Kramer’s ions---the degeneracy can be lifted completely. The species with an odd number of electrons---the Kramer’s ions---require an external field to fully break this degeneracy. Rare-earth ions can have further splittings of their crystal field levels from hyperfine structure, the Zeeman effect, and higher order moments \cite{thiel2011rare, liu2006spectroscopic}.  

While these electronic levels are largely insensitive to their environment, it is residual environmental effects that broaden the optical and spin transitions. Homogeneous broadening is largely due to dynamical processes within the solid and can be sorted broadly into four categories: ion-phonon interactions, ion-ion interactions, ion-host lattice interactions, and pure dephasing. Optical phonons are suppressed at or below 4 K \cite{liu2006spectroscopic,thiel2011rare, kunkel2018Recent}. Microwave phonons can couple the spin transitions and primarily interact with the ion via multiphonon processes that scale nonlinearly with temperature \cite{liu2006spectroscopic,arcangeli2015temperature,hastings2008zeeman}. Ion-ion interactions account for a number of interactions such as instantaneous spectral diffusion, in which the excitation of one ion shifts the electric dipole of a neighboring ion \cite{liu2006spectroscopic, yen1965Optical,thiel2011rare, kunkel2018Recent, kinos2022Microscopic}. The dominant source of dephasing below 4~K is interactions between the ion and the host lattice, which acts as a spin bath \cite{thiel2011rare}. For example, fluctuating magnetic fields due to nuclear spin flips of Y$^{3+}$ are the dominant source of dephasing in YSO \cite{zhou2018probing}. This dephasing can be suppressed to extend the coherence times by application of a moderate external magnetic field to polarize the spin bath, as well as finding spin transitions that have zero effective first-order Zeeman shift \cite{fraval2004method, longdell2006characterization, mcauslan2012reducing, zhong2015optically}, and employing rephasing techniques like dynamical decoupling \cite{beavan2009demonstration, hain2022few,pascual2012spin,lovric2013Faithful, fraval2005Dynamic}. 

Defects and strain in the crystalline host result in inhomogeneous broadening of both the optical and spin transitions. The local crystal field seen by the ions varies site-to-site, resulting in a static inhomogeneous shift to their transition energy. In commercially doped rare-earth materials, the inhomogeneous broadening of the optical transition can be on the order of 10's of MHz to 100’s of GHz depending on factors such as (but not limited to) the host material, doping concentration, and isotopic purity. The inhomogeneous broadening is always many orders of magnitude larger than the homogeneous linewidth \cite{macfarlane1992inhomogeneous, lafitte2022optical, thiel2010optical,  thiel2014tm, tiranov2018spectroscopic, sun2012optical, liu2006spectroscopic, thiel2011rare,kunkel2018Recent,sellars2004Investigation}. In the case of rare-earth doped solids, where the ions themselves are substitutional dopants, they may be the primary source of this inhomogeneous broadening \cite{ahlefeldt2013Evaluation,ahlefeldt2016ultranarrow,thiel2011rare,campbell2019echo}.

Ions with an even number of electrons---the non-Kramer’s ions (Eu$^{3+}$, Pr$^{3+}$, Tm$^{3+}$)---have a quenched electronic spin and consequently are more insensitive to their environment and have longer coherence properties. The nuclear spin state splittings in these ions are between 10-100 MHz---too narrow to be optically resolved in many host materials due to the large inhomogeneous linewidth \cite{equall1995homogeneous,sun2012optical,karlsson2017Nuclear,goldner2004Magnetic,nakamura2014Spectroscopy,louchet2008Optical,longdell2006characterization,fan2019electromagnetically}. The spin coherence properties of the non-Kramer’s ions can be exploited for long storage times by utilizing techniques such as spectral holeburning to resolve the nuclear spin states; however, this comes at the expense of sacrificing optical depth---and therefore the storage efficiency---by only using a small subset of the atoms in the inhomogeneous line \cite{campbell2019echo}. 

In the Kramer’s ions (Er$^{3+}$,Yb$^{3+}$, Nd$^{3+}$)---the paramagnetic species---application of an external magnetic field can yield Zeeman splittings large enough for the Zeeman sublevels to to be resolvable in the inhomogeneous line. These species are of particular interest due to their larger bandwidths, on the order of hundreds of MHz to GHz, and potential for interfacing with platforms such as superconducting qubits \cite{kindem2018Characterization,probst2015Microwave,ortu2018Simultaneous,welinski2019Electron}.  While their Zeeman states can be accessible and have the potential for larger memory bandwidths than the non-Kramer’s ions, with state splittings of several hundred MHz to GHz, their electronic spin has stronger coupling to the environment due to its larger magnetic moment and thus is more susceptible to fluctuations and exhibits shorter coherence times. A challenge of utilizing the Kramers species is the ability to efficiently optically pump them \cite{cruzeiro2018Efficient,cruzeiro2017Spectral}, which is a prerequisite of many quantum memory protocols such as AFC and CRIB \cite{afzelius2009multimode,lvovsky2009optical,campbell2019echo}. Erbium is one of the most attractive rare-earths due to its telecom C-band transition at 1.5 $\mu$m; however erbium can be difficult to optically pump since the lifetime of its Zeeman sublevels is only about an order of magnitude slower than the excited state lifetime \cite{hashimoto2016Coherent, hastings2008zeeman}. However, using Kramers isotopes with nuclear spin helps substantially and spin coherence times as long as 1 second have been achieved in $^{167}$Er$^{3+}$:YSO \cite{ranvcic2018coherence}. Progress has been made recently to overcome the limitations of optical pumping in both Yb$^{3+}$ and Nd$^{3+}$ \cite{cruzeiro2018Efficient,welinski2020Coherence,zhong2017interfacing}. 

 There have been many proof-of-principle light storage experiments with both classical light and weak coherent states that show these systems can operate with the proper SNR to store single photons \cite{timoney2013single, kutluer2016spectral,stuart2021Initialization,zhong2017Nanophotonic}. Storing single-photon states in a rare-earth-ion-based quantum memory has also been achieved \cite{saglamyurek2011broadband,seri2017quantum,bussieres2014quantum, rielander2014Quantum, zhu2022demand, seri2018laser, seri2019quantum, gundougan2012quantum, liu2021heralded, clausen2011Quantum, rakonjac2021Entanglement,lago2022long,lago2021telecom}.
 
 To date, the longest storage time of a bright classical pulse was 53 minutes and for weak coherent states exceeding 1.0 second \cite{ma2021one, hain2022few}; however, both were acheived with low storage efficiencies. On-chip storage of telecommunications C-band light at the few-photon level has also been demonstrated in a photonic nanoresonator with isotopically purified Er-167:YSO \cite{craicui2019nanophotonic}. Rare-earth ions have demonstrated quantum memory protocols with bandwidths exceeding a GHz \cite{saglamyurek2011broadband,liu2021heralded,askarani2019storage,davidson2020improved}.

\section{Theory}\label{theory_sec} 

\subsection{Maxwell-Bloch Equations}\label{MB_sec}

In this section, we expand upon the three-level Maxwell-Bloch equations described in Sec.~\ref{ensemble_sec}. Section~\ref{hom_MB_sec} provides several approximations to the Maxwell-Bloch equations relevant for  the memory protocols discussed in Sec.~\ref{protocol_sec} that rely on homogeneous broadening, while Sec.~\ref{inhom_MB_sec} provides more information on the Maxwell-Bloch equations in the presence of inhomogeneous broadening.

Each of the protocols in Sec.~\ref{protocol_sec} requires or permits a certain type of line broadening for the intermediate excited state shown in Fig.~\ref{fig_levelstruc}, and the type of broadening present in a given system limits the possible memory protocols that can be implemented in that system. The types of excited-state line broadening are broadly categorized into those with homogeneous and inhomogeneous mechanisms. Homogeneous broadening mechanisms affect each atom in the system equally, whereas inhomogeneous mechanisms affect different classes of atoms differently. These classes of atoms may be distinguished in frequency, spatial position, momentum, or any other degree of freedom. Homogeneous broadening mechanisms, which typically apply to cold atoms---but, importantly for broadband operation, are not restricted to cold atoms---include effects such as natural or lifetime broadening, resonance broadening, collisional broadening (which is separated into impact and quasi-static regimes, depending on the timescale of collisions \cite{corney1978atomic}), and phonon broadening in solids. Inhomogeneous broadening mechanisms include Zeeman and Stark broadening (these are typically considered inhomogeneous due to spatial or temporal non-uniformity in the magnetic and electric fields, respectively), Doppler broadening, and crystal-field or impurity-based broadening in solids \cite{liu2006spectroscopic}.

\subsubsection{Homogeneous}\label{hom_MB_sec}

Several approximations to Eqs.~\eqref{Aeq_t_hom}-\eqref{Beq_t_hom} from Sec.~\ref{ensemble_sec} lead to the mathematical descriptions of EIT, ATS, ATT, SR, and Raman protocols, which we compile here. In addition, we provide representations of the Maxwell-Bloch equations in the spectral domain and in the temporal domain split into amplitude and phase. These different forms of the Maxwell-Bloch equations are useful for both analytical and numerical investigation of quantum memory behavior.

In the case of EIT, the approximations $\Delta=0$ (i.e., $\bar{\gamma}=1$) and $BW\ll\gamma$ (or $\tFWHM\gamma\gg1$) for both the signal and control fields implies that the time evolution of the atomic polarization in Eq.~\eqref{Peq_t_hom} may be adiabatically eliminated, and a reduced set of partial differential equations can be found:

\begin{align}
    \label{Aeq_t_EIT}\partial_z A(z,\tau) &= -dA(z,\tau) +i\sqrt{d}\frac{\Omega(\tau)}{2}B(z,\tau)\\
    \label{Beq_t_EIT}\partial_\tau B(z,\tau) &= -i\sqrt{d}\frac{\Omega^*(\tau)}{2} A(z,\tau) - \frac{\abs{\Omega(\tau)}^2}{4}B(z,\tau),
\end{align}

\noindent which are significantly less computationally expensive to simulate. 

In the case of ATS, the assumptions $\Delta=0$, $\abs{\Omega}\gg\bar{\gamma},\sqrt{d}$ imply the elimination of the first two terms on the right-hand-side of Eq.~\eqref{Peq_t_hom}, and in the case of a constant control field (which without loss of generality can be assumed to be purely imaginary), the simplified coupled partial differential equations

\begin{align}
    \label{Peq_t_ATS}\partial_\tau P(z,\tau) &= - i\frac{\Omega(\tau)}{2} B(z,\tau)\\
    \label{Beq_t_ATS}\partial_\tau B(z,\tau) &= -i\frac{\Omega^*(\tau)}{2} P(z,\tau),
\end{align}

\noindent lead to periodically oscillating atomic polarization and spin wave populations that are $\pi/2$ out of phase:

\begin{align}
    \label{Peq_t_ATS2}A(z,\tau) \propto P(z,\tau) &= \sin\left(\frac{\abs{\Omega(\tau)}}{2}t\right)\\
    \label{Beq_t_ATS2}B(z,\tau) &= \cos\left(\frac{\abs{\Omega(\tau)}}{2}t\right).
\end{align}

For ATT, the assumptions $\Delta=0$, $BW\geq\gamma$ (or $\tFWHM\gamma\leq1$), and $d\gg1$ imply that we can drop the $\bar{\gamma}P(z,\tau)$ term in Eq.~\eqref{Peq_t_hom}. Further assuming a $\pi$-pulse control field that arrives after absorption of the signal field implies the terms involving $\Omega(\tau)$ are eliminated during the initial absorption stage, leaving only the simplified equations:

\begin{align}
    \label{Aeq_t_ATT}\partial_z A(z,\tau) &= - \sqrt{d} P(z,\tau)\\
    \label{Peq_t_ATT}\partial_\tau P(z,\tau) &= \sqrt{d} A(z,\tau),
\end{align}

\noindent which has exact solution 

\begin{align}
    \label{Aeq_t_ATT2}A(z,\tau) &= e^{t-dz}[1-\Theta(\Deltau)]\\
    \label{Peq_t_ATT2}P(z,\tau) &= \sqrt{d}e^{t-dz}[1-\Theta(\Deltau)]
\end{align}

\noindent during the absorption stage (until the arrival of the control field, $\tau<\Deltau$), where $\Theta(\Deltau)$ is the Heaviside step function. The $\pi$-pulse control field then in principle transfers $P\rightarrow B$ exactly. Applying the definition of storage efficiency, we find $\eta_\text{stor} = 1-e^{-2d}$. The treatment of the SR memory protocol is nearly identical, but operates under the assumption of $BW\gg\gamma$ (or $\tFWHM\gamma\ll1$) and superradiant coupling between atoms in the ensemble. The approximations leading to the Raman protocol are similar to EIT, but in the far-off-resonant limit, i.e., $\Delta\gg\gamma$ (i.e., $\bar{\gamma}\approx-i\Delta/\gamma = -i\bar{\Delta}$) and $BW\ll\Delta$. These conditions also imply the adiabatic elimination of the atomic polarization, leading to the simplified system:

\begin{align}
    \label{Aeq_t_Raman}\partial_z A(z,\tau) &= -i\frac{d}{\bar{\Delta}}A(z,\tau) -\frac{\sqrt{d}}{\bar{\Delta}}\frac{\Omega(\tau)}{2}B(z,\tau)\\
    \label{Beq_t_Raman}\partial_\tau B(z,\tau) &= \frac{\sqrt{d}}{\bar{\Delta}}\frac{\Omega^*(\tau)}{2} A(z,\tau) - i\frac{\abs{\Omega(\tau)}^2}{4\bar{\Delta}}B(z,\tau).
\end{align}

Several other forms of the Maxwell-Bloch equations are useful both analytically and numerically. In the spectral domain, we can rewrite the Maxwell-Bloch equations via application of the Fourier transform as:

\begin{align}
	\label{A_PDE_FT}\partial_z \tilde{A}(z,\omega) &= -\sqrt{d}\tilde{P}(z,\omega)\\
	\label{P_PDE_FT} i\omega \tilde{P}(z,\omega) &= -\bar{\gamma}\tilde{P}(z,\omega) + \sqrt{d}\tilde{A}(z,\omega) - \frac{i}{\sqrt{2\pi}}\frac{\tilde{\Omega}(\omega)}{2}*\tilde{B}(z,\omega)\\
	\label{B_PDE_FT}i\omega \tilde{B}(z,\omega) &= -\gamma_B \tilde{B}(z,\omega) - \frac{i}{\sqrt{2\pi}}\frac{\tilde{\Omega}^*(-\omega)}{2}*\tilde{P}(z,\omega).
\end{align}

\noindent Here we have made use of the convolution theorem for Fourier transforms and the fact that the Fourier transform of a conjugated function is the conjugate of the function's Fourier transform reflected about $\omega=0$. We have used ``$*$" to express the convolution of two functions, e.g. $\tilde{\Omega}(\omega)*\tilde{B}(z,\omega) = \int_{-\infty}^{\infty}du\, \tilde{\Omega}(\omega-u)\tilde{B}(z,u)$. Here $\omega$ represents the difference from each field's center frequency. This form of the Maxwell-Bloch equations can be particularly useful for simulation of broadband quantum memory. As usual, application of the Fourier transform has converted equations involving temporal derivatives to simple algebraic equations.

Another useful form of the Maxwell-Bloch equations arises from separating the amplitude and phase of each field, leading to the following 6 equations of motion:

\begin{align}
    \label{Aeq_t_amp}\partial_z \abs{A(z,\tau)} &= -\sqrt{d} \abs{P(z,\tau)}\cos[\phi_P(z,\tau) - \phi_A(z,\tau)]\\
    \partial_z \phi_A(z,\tau) &= -\sqrt{d}\frac{\abs{P(z,\tau)}}{\abs{A(z,\tau)}}\sin[\phi_P(z,\tau) - \phi_A(z,\tau)]\\
    \notag \partial_\tau \abs{P(z,\tau)} &= -\bar{\gamma} \abs{P(z,\tau)} + \sqrt{d} \abs{A(z,\tau)}\cos[\phi_A(z,\tau)-\phi_P(z,\tau)]\\
    \label{Peq_t_amp} &\hspace{2em} +\frac{1}{2}\Re[\Omega(\tau)]\abs{B(z,\tau)}\sin[\phi_B(z,\tau)-\phi_P(z,\tau)]\\
    \notag &\hspace{3em} + \frac{1}{2}\Im[\Omega(\tau)]\abs{B(z,\tau)}\cos[\phi_B(z,\tau)-\phi_P(z,\tau)]\\
    \notag \partial_\tau \phi_P(z,\tau) &= \Delta + \sqrt{d} \frac{\abs{A(z,\tau)}}{\abs{P(z,\tau)}}\sin[\phi_A(z,\tau)-\phi_P(z,\tau)] \hspace{0em}\\
    \label{Peq_t_phase} &\hspace{2em} -\frac{1}{2}\Re[\Omega(\tau)]\frac{\abs{B(z,\tau)}}{\abs{P(z,\tau)}}\cos[\phi_B(z,\tau)-\phi_P(z,\tau)]\\
    \notag &\hspace{3em} + \frac{1}{2}\Im[\Omega(\tau)]\frac{\abs{B(z,\tau)}}{\abs{P(z,\tau)}}\sin[\phi_B(z,\tau)-\phi_P(z,\tau)]\\
    \notag\partial_\tau \abs{B(z,\tau)} &= -\gamma_B \abs{B(z,\tau)} + \frac{1}{2}\Re[\Omega(\tau)]\abs{P(z,\tau)}\sin[\phi_P(z,\tau)-\phi_B(z,\tau)]\\
    \label{Beq_t_amp} &\hspace{3em}- \frac{1}{2}\Im[\Omega(\tau)]\abs{P(z,\tau)}\cos[\phi_P(z,\tau)-\phi_B(z,\tau)]\\
    \notag\partial_\tau \phi_B(z,\tau) &= - \frac{1}{2}\Re[\Omega(\tau)]\frac{\abs{P(z,\tau)}}{\abs{B(z,\tau)}}\cos[\phi_P(z,\tau)-\phi_B(z,\tau)]\\
    \label{Beq_t_phase} &\hspace{3em}- \frac{1}{2}\Im[\Omega(\tau)]\frac{\abs{P(z,\tau)}}{\abs{B(z,\tau)}}\sin[\phi_P(z,\tau)-\phi_B(z,\tau)],
\end{align}

\noindent where each field is written $X(z,\tau) = \abs{X(z,\tau)}e^{i\phi_X(z,\tau)}$ for $X=A,P,$ and $B$, and where $\Re(\cdot)$ and $\Im(\cdot)$ designate the real and imaginary parts of the enclosed function, respectively. This form of the equations of motion is particularly useful for analysis and simulations involving chirped optical fields or fields with non-trivial temporal/spectral phase.

\subsubsection{Inhomogeneous}\label{inhom_MB_sec}

Inhomogeneous broadening of the intermediate excited state is treated as a coherent sum of homogeneous distributions separated in frequency \cite{gorshkov2007photon_3}, leading to the following equations of motion:

\begin{align}
    \label{Aeq_inhom}\partial_z A(z,\tau) &= -\sqrt{d} P(z,\tau)\\
    \label{Peq_inhom}\partial_\tau P_\Delta(z,\tau) &= -(\bar{\gamma}-i\Delta) P_\Delta(z,\tau) + \sqrt{d} \sqrt{p_\Delta}A(z,\tau) - i\frac{\Omega(\tau)}{2} B_\Delta(z,\tau)\\
    \label{Beq_inhom}\partial_\tau B_\Delta(z,\tau) &= -\gamma_B
    B_\Delta(z,\tau) -i\frac{\Omega^*(\tau)}{2} P_\Delta(z,\tau),
\end{align}

\noindent where $P(z,\tau)=\int d\Delta \sqrt{p_\Delta} P_\Delta(z,\tau)$ in Eq.~\eqref{Aeq_inhom} and $p_\Delta$ represents the normalized fraction of atoms with inhomogeneous detuning $\Delta$ (i.e., $\int d\Delta \, p_\Delta = 1$). Here we have redefined $\bar{\gamma} = (\gamma - i\Delta_0)/\gamma$ to designate the detuning $\Delta_0$ of the optical fields relative to the atoms with no inhomogeneous shift (i.e., for $\Delta=0$).

Analytic expressions for memory efficiency and other metrics for each of the protocols discussed in Sec.~\ref{protocol_sec} may be derived from Eqs.~\eqref{Aeq_inhom}-\eqref{Beq_inhom} by assuming the appropriate inhomogeneous spectral profile $p_\Delta$ and the appropriate pulse sequence [i.e., for most protocols, during storage the control field is off and terms involving $\Omega(\tau)$ in Eqs.~\eqref{Aeq_inhom}-\eqref{Beq_inhom} may be neglected].

Again, these equations may be Fourier transformed into the spectral domain or split into amplitude and phase for ease of computation, just as in the homogeneous case described above.

\subsection{Efficiency Optimization}

The primary focus of most existing theoretical work dealing with atomic ensemble quantum memory revolves around the goal of increasing memory efficiency. In principle, the same or similar tools that have been developed for efficiency optimization may also be used to improve other metrics, but efficiency is uniquely important for quantum memory applications and has accordingly received the majority of theoretical attention. The techniques described in the following Secs.~\ref{sigShaping_sec} and \ref{ctrlShaping_sec} are most fully described in Refs.~\cite{gorshkov2007photon_1,gorshkov2007photon_2,gorshkov2007photon_3,gorshkov2007universal,nunn2008quantum}.

\subsubsection{Signal Field Shaping}\label{sigShaping_sec}

One particularly powerful tool for increasing memory efficiency is optimization of the temporal profile of the signal field in both amplitude and phase. The idea behind this optimization technique is that every stage of the memory process can be treated as a linear integral mapping with a unique integral kernel fully describing the map. Each kernel can be decomposed into singular vectors and associated singular values that act as a prefactor describing the efficiency for each mapping between singular vectors. For example, the storage process is described by the map

\begin{equation}
    \label{Bout_eq}B_\text{out}(z) = \int_{-\infty}^{\infty} d\tau\, K_\text{stor}(z,\tau)A_\text{in}(\tau),
\end{equation}

\noindent where $A_\text{in}(\tau)$ is the incident signal field temporal profile, $K_\text{stor}(z,\tau)$ is the integral kernel for the storage operation, and $B_\text{out}(z)$ is the long-lived spatially dependent spin wave resulting from the storage process. Singular value decomposition of this storage kernel takes the form

\begin{equation}
    \label{K_eq}K_\text{stor}(z,\tau) = \sum_j \lambda_j B_j(z) A^*_j(\tau),
\end{equation}

\noindent where $B_j(z)$ and $A_j(\tau)$ are the left-singular and right-singular vectors (sometimes called the `optimal modes') of $K_\text{stor}(z,\tau)$ and $\lambda_j$ are its singular values. The exact form of the storage kernel $K_\text{stor}(z,\tau)$ depends on parameters of the memory such as the optical depth, linewidth, control field shape, etc. Given Eqs.~\eqref{Bout_eq} and \eqref{K_eq}, it is clear to see that the largest storage efficiency is achieved when $A_\text{in}(\tau)$ matches the right-singular vector of the storage kernel with the largest singular value, $A_\text{opt}(\tau)$, as any admixture of $A_\text{in}(\tau)$ with other modes will imply a component of the mapping with non-optimal $\lambda_j$, or non-optimal efficiency. Thus if it is possible to arbitrarily shape the signal field to be stored, one can optimize memory efficiency by shaping $A_\text{in}(\tau)$ to match $A_\text{opt}(\tau)$. One can determine the exact form of $A_\text{opt}(\tau)$ by numerically or analytically constructing $K_\text{stor}(z,\tau)$ and computing its singular value decomposition based on the experimental parameters at hand, or by an iterative experimental process based on time-reversal described in Refs.~\cite{novikova2012electromagnetically,gorshkov2007photon_2,phillips2008optimal,novikova2007optimal}.

\subsubsection{Control Field Shaping}\label{ctrlShaping_sec}

Shaping of incident single-photon light pulses that contain sensitive encoded quantum information can introduce loss, which is undesirable for almost every application of quantum memory. It is generally more desirable to shape the strong, many-photon control field, where loss can be easily compensated for by increasing control field power. The approach for optimizing memory efficiency through the temporal shape of the control field then takes on a similar form to the signal-field shaping discussed above. In the case of quantum storage, instead of matching the signal field to the right-singular vector of the storage kernel with the largest singular value, the task becomes to shape the storage kernel itself and thereby shape its right-singular vector with largest eigenvalue to match the input signal field. This can be accomplished analytically in several limiting cases, or, in the general case, this can be accomplished numerically through an iterative process: First, the storage kernel is constructed numerically for an initial guess of the control field shape (amplitude and phase), $\Omega_1(\tau)$, which is discretized along a series of spline points. The optimal mode $A^1_\text{opt}(\tau)$ of this initial storage kernel is constructed and compared to the target signal field $A_\text{in}(\tau)$. A series of $N$ numeric interpolations between $A^1_\text{opt}(\tau)$ and $A_\text{in}(\tau)$ are constructed, $A_k(\tau)$, where $A_1(\tau)=A^1_\text{opt}(\tau)$ and $A_N(\tau)=A_\text{in}(\tau)$. For each $A_k(\tau)$, the control field spline points are optimized, leading to an optimal control field shape $\Omega_k(\tau)$ that maximizes storage efficiency for each $A_k(\tau)$ and provides the initial guess for $\Omega_{k+1}(\tau)$. This optimization at each step can either be implemented by constructing the storage kernel for each guess and comparing the optimal mode from the SVD of the kernel with $A_k(\tau)$, or by numerically integrating the single instance of the Maxwell-Bloch equations with input $A_k(\tau)$ and the control field guess. Assuming a large enough number of interpolations $N$, this iterative process slowly transforms the storage kernel such that its optimal mode overlaps with the final $A_N(\tau)$ (i.e., $A^N_\text{opt}(\tau) = A_N(\tau) = A_\text{in}(\tau)$, where the superscript $N$ is just a label, not a multiplicative power).

In practice, this optimization technique tends to produce similar efficiencies compared to the signal-field shaping technique, however the relative strength of each technique for optimizing memory efficiency remains an open question. It is possible that regions of the memory parameter space exist where control field shaping either outperforms or underperforms signal-field shaping, and it is possible that optimization of both fields together may yield higher efficiencies in some regions than optimization of either field shape alone.

\begin{figure}[t]
	\centering
	\includegraphics[width=1\linewidth]{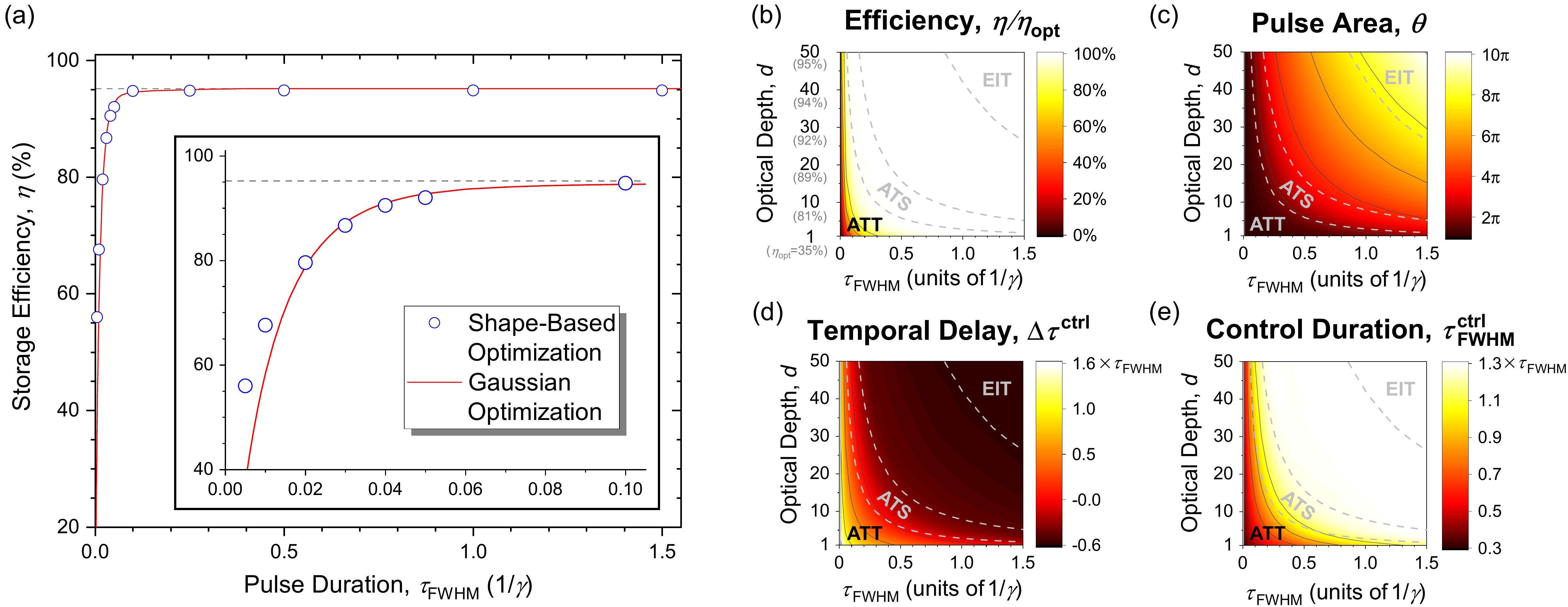}
	\caption{(a) Comparison of signal-field shape-based and Gaussian-pulse-shape optimization techniques for an optical depth of $d=50$. The dotted line marks the optimal storage efficiency, $\eta_\text{opt}\approx95\%$. (b) Gaussian-optimized memory efficiency and (c)-(e) optimized control field parameters in the broadband regime. Adapted from Ref.~\cite{shinbrough2021optimization}.}
	\label{fig_gaussian_opt}
\end{figure}

\subsubsection{Gaussian-Pulse-Shape Optimization}\label{GaussOpt_sec}

In contrast to the two techniques described above, each of which requires arbitrary shaping of the respective light field temporal envelope, recent work reported in Ref.~\cite{shinbrough2021optimization} addresses the question of how far memory efficiency can be optimized using a fixed shape of the temporal envelope, one which is a good approximation for the shape that is natively produced by modelocked lasers and single-photon sources. The shape chosen in Ref.~\cite{shinbrough2021optimization} is a Gaussian, such that the signal and control field shapes are

\begin{align}
    A_\text{in}(\tau) &= e^{-\tau^2/4\sigma^2}\\
    \Omega(\tau) &= \Omega_0\, e^{-[(\tau-\Deltau)/2\sigma^\text{ctrl}]^2},
\end{align}

\noindent where again all timescales are normalized by $1/\gamma$, and $\sigma=\tFWHM/(2\sqrt{2\ln{2}})$, $\Omega_0 = \theta/(2\sqrt{\pi}\sigma^\text{ctrl})$, the control-field pulse area is $\theta = \int_{-\infty}^\infty d\tau\, \Omega(\tau)$, the temporal delay of the control field relative to the maximum of the signal field is $\Deltau$, and the duration of the control field is $\tctrl=2\sqrt{2\ln{2}}\sigma^\text{ctrl}$.

The approach of Gaussian-pulse-shape optimization has three main benefits: First, related to the motivation of using the pulse shapes that are native to many modelocked lasers and single-photon sources, it demonstrates that over a large range of memory parameter space (for a large range of optical depths $d$ and signal durations/linewidths $\tFWHM\gamma$) very high storage efficiencies very close to the optimal bound discussed in Sec.~\ref{metric_sec} can be achieved with only these native Gaussian shaped pulses. This eliminates the need in most quantum memories for experimentally complex and costly pulse shaping methods. Fig.~\ref{fig_gaussian_opt}(a) shows a comparison of the storage efficiencies achieved using both signal-field, arbitrary shape-based optimization and the optimization of Gaussian-shaped control fields for $d=50$. Over a wide range of signal-field pulse durations $\tFWHM\gamma$, the Gaussian-pulse-shape optimization technique produces the same efficiency as the shape-based technique, within numerical error. For the most broadband photons (smallest $\tFWHM\gamma$), the shape-based technique outperforms Gaussian-pulse-shape optimization, but the storage efficiency is far from the optimal bound in both cases. Additionally, the Gaussian-pulse-shape optimization technique is significantly less computationally expensive, allowing for a continuous calculation of optimized storage efficiency as a function of $\tFWHM\gamma$, whereas the shape-based optimization method is only computationally tractable in a point-wise fashion. 

Second, Gaussian-pulse-shape optimization provides three physically intuitive optimization parameters for the control field --- the control-field pulse area, duration, and delay relative to the signal field --- whereas the shape-based technique relies on arbitrary pulse shaping that is arguably less physically intuitive. The resulting optimized control field parameters are shown in Fig.~\ref{fig_gaussian_opt}(c)-(e). In addition to providing a convenient lookup table for the optimized control parameters to use experimentally for a quantum memory with a given optical depth, signal duration, and linewidth, Fig.~\ref{fig_gaussian_opt}(c)-(e) also agrees with and builds upon the physical intuition for each quantum memory protocol. In the EIT region, Fig.~\ref{fig_gaussian_opt}(c)-(e) shows that control fields with large duration and pulse area should be used that arrive before the signal field; control fields of this type open the well-known transparency window of EIT and adiabatically close this window as the signal field enters the medium and is compressed and stored. Similarly the physical intuition holds in the ATS and ATT regions, where control fields of $2\pi$ and $\pi$ pulse area, control durations of $\tFWHM$ and less than $\tFWHM$, and delays of 0 and positive delays implying a control field that arrives after the signal field are used, respectively, as predicted for both of these protocols. These physically intuitive control field parameters are practically useful when dealing with tunable experiments, but also allow for more sophisticated sensitivity analysis, as detailed in Ref.~\cite{shinbrough2022variance} and Sec.~\ref{sensitivity_sec}.

Third, Gaussian-pulse-shape optimization reveals a continuous mathematical transformation between the three resonant protocols of EIT, ATS, and ATT. As shown in Fig.~\ref{fig_gaussian_opt}(b) and (c)-(e), memory efficiency remains at the optimal bound across a continuous range of memory parameters $\mathcal{M} = (d,\tFWHM\gamma)$ where the memory protocol behavior changes continuously from EIT to ATS to ATT. As $\mathcal{M}$ varies, the control field parameters $\mathcal{G}=\left(\theta,\Deltau,\tctrl\right)$ also vary continuously and monotonically. Na{\"i}vely, one might predict discontinuities in $\mathcal{G}$ as one protocol region changes into another, but Fig.~\ref{fig_gaussian_opt} shows this is not the case, and a continuous transformation between memory protocols exists and can be exploited to achieve optimal memory efficiency in the regions of memory parameter space between protocols, or in the regions of ``mixed'' memory behavior.

\subsubsection{Inhomogeneous Profile Shaping}

While in principle the three optimization techniques above (Sec.~\ref{sigShaping_sec}-\ref{GaussOpt_sec}) can be applied to both homogeneous and inhomogeneous systems, they have primarily been applied in the homogeneous case. In the inhomogeneous case another unique degree of freedom is available for optimization, namely the inhomogeneous broadening profile, which can be straightforwardly manipulated via spectral hole burning and related experimental techniques.

In Ref.~\cite{gorshkov2007photon_3}, the authors consider several inhomogeneous profile shapes, including a delta function (identical to the homogeneous case), a Lorentzian distributed inhomogeneous profile, and a Gaussian distributed inhomogeneous profile typical of Doppler-broadened gases. The efficiencies resulting from the three inhomogeneous are compared (where, unsurprisingly, the delta function profile leads to the highest efficiency), and a general procedure for calculating the efficiency for different inhomogeneous profile shapes is developed. In Ref.~\cite{bonarota2010efficiency}, a similar analysis is carried out assuming a periodic inhomogeneous frequency profile characteristic of the AFC protocol. The authors show that in this case, the optimal inhomogeneous profile shape is a square-toothed frequency comb with an analytically calculable tooth width depending on the achievable peak optical depth. In addition to these techniques, one can imagine implementing a similar procedure to Sec.~\ref{GaussOpt_sec} to calculate the optimal inhomogeneous profile shape in the absence of the assumptions made in the AFC protocol. Ref.~\cite{fan2019electromagnetically}, for example, has examined the EIT visibilities resulting from different inhomogeneous shapes, which may lead to the discovery of an optimal inhomogeneous shape for the EIT protocol in inhomogeneously broadened media.

\subsection{Sensitivity Analysis}\label{sensitivity_sec}

Although efficiency optimization has received the lion's share of theoretical attention related to atomic ensemble quantum memory, recent work reported in Ref.~\cite{shinbrough2022variance} introduces theoretical tools aimed at the distinct task of characterizing and comparing the \textit{sensitivity} of different quantum memory protocols to external perturbations.

\begin{figure}[t]
	\centering
	\includegraphics[width=1\linewidth]{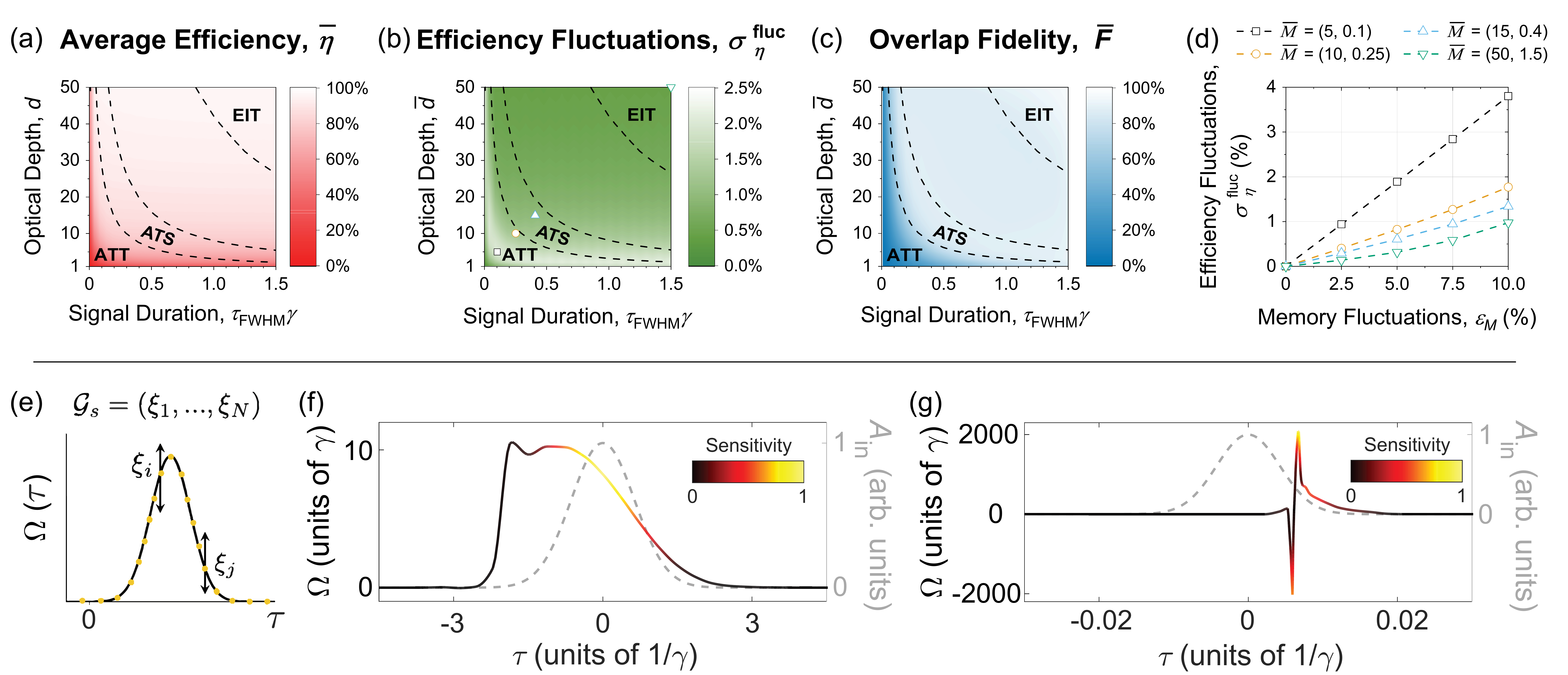}
	\caption{Select results from sensitivity analysis of $\Lambda$-type quantum memory in the presence of (a)-(d) short-timescale fluctuations of the memory parameters (optical depth, $d$, and linewidth, $\gamma$), and (e)-(g) long-timescale drift in control field parameters.}
	\label{fig_sensitivity}
\end{figure}

Figure \ref{fig_sensitivity} shows a few selected results from Ref.~\cite{shinbrough2022variance}, considering the case of short-timescale, shot-to-shot fluctuations $\zeta = (\zeta_d,\zeta_g)$ in memory parameters $\mathcal{M}$ drawn from a normal distribution $P(\zeta) \sim e^{-(\zeta_d^2 g^2 + \zeta_g^2 d^2)/[2(\epsilon_M dg)^2]}$, where $g=\tFWHM\gamma$ [Fig.~\ref{fig_sensitivity}(a)-(d)], and long-timescale drift in control field parameters $\mathcal{G}_s$  for a control field with arbitrary profile defined by spline points $\mathcal{G}_s = (\xi_1,...,\xi_N)$ [Fig.~\ref{fig_sensitivity}(e)-(g)]. The situation in mind addressed by Fig.~\ref{fig_sensitivity}(a)-(d) is frequently encountered in atomic-ensemble quantum memory using hot atoms, wherein the average optical depth and excited state linewidth are fixed with minimal long-timescale drift (typically through setting the vapor cell temperature) but there may be non-negligible shot-to-shot fluctuations in optical depth and linewidth arising from density fluctuations in the interaction volume defined by the optical modes. With this situation in mind, Fig.~\ref{fig_sensitivity}(a) shows average memory efficiency in the presence of memory parameter fluctuations $\epsilon_M = 5\%$, and Fig.~\ref{fig_sensitivity}(b) shows the magnitude of the fluctuations in memory efficiency arising from $\epsilon_M = 5\%$. First, the average memory efficiency drops slightly from the case without fluctuations [e.g., compared to Fig.~\ref{fig_gaussian_opt}(b)], as expected. Second, we find that the absorb-then-transfer (ATT) region is most sensitive to these memory parameter fluctuations, as it is this region that possesses the largest fluctuations in memory efficiency (for the same fluctuations in memory parameters) shown in Fig.~\ref{fig_sensitivity}(b). The relatively large sensitivity of the ATT protocol compared to the other resonant memory protocols (i.e., ATS and EIT) is explained by the overlap fidelity of optimal control fields neighboring any given point in $\mathcal{M}=(d,\tFWHM\gamma)$, shown in Fig.~\ref{fig_sensitivity}(c). Where the overlap fidelity is low, one expects a significant mismatch between the control field used (which corresponds to the average optical depth and linewidth) and the optimal control field (corresponding to the optical depth and linewidth modified by a given shot-to-shot fluctuation), which should result in lower average efficiency and larger efficiency fluctuations. When the overlap fidelity is high, one expects (and we observe) high average efficiency and lower efficiency fluctuations. Across several points in the memory parameter space shown by the markers in Fig.~\ref{fig_sensitivity}(b), sampling over all three resonant memory protocols, we measure efficiency fluctuations as a function of the memory parameter fluctuation magnitude, $\epsilon_M$, and plot the results in Fig.~\ref{fig_sensitivity}(d). First, we observe efficiency fluctuations that are smaller, as a percentage, than the memory parameter fluctuations that cause them (i.e., memory parameter fluctuations of magnitude $\epsilon_M = 5\%$ cause at most memory efficiency fluctuations of 2.5\%). This behavior persists over all reasonable magnitudes of memory parameter fluctuations, shown in Fig.~\ref{fig_sensitivity}(d), which implies that all three resonant memory protocols are `stable.' We also note that the EIT and ATS protocols are significantly less sensitive to memory parameter fluctuations than the ATT protocol (by about a factor of 3), across all memory parameter fluctuation magnitudes.

In addition to deviations in memory parameters, we can also consider the effect of deviations in control field parameters within the same general mathematical framework. Typically memory parameters fluctuate over a short timescale around a fixed center point; control field parameters may also fluctuate in this manner, but a more common problem experimentally is long timescale drift from an initially optimal setpoint. Investigating the sensitivity of a memory to this long timescale drift is, coincidentally, equivalent to probing the difficulty of finding the optimal setpoint for that memory. Memory protocols that allow for a larger region of control field parameter space with optimal efficiency, and which therefore lead to less difficulty in finding the optimal control field setpoint, are less sensitive to control field drift out of that optimal region. The converse is also true---memory protocols that permit only a small region of control field parameter space with optimal efficiency lead to more difficulty in finding the optimal control field setpoint and are more sensitive to control field drift. This situation is explored in Fig.~\ref{fig_sensitivity}(e)-(g) for optimal arbitrary control field shapes. Fig.~\ref{fig_sensitivity}(e) shows the general approach; the optimal control field shape for a certain memory protocol is found by the procedure described in Sec.~\ref{ctrlShaping_sec}, and each spline point parameterizing the control field shape is allowed to drift by $\epsilon_G = 5\%$ of its optimum. The resulting change in memory efficiency is recorded and is used to compute a normalized sensitivity relative to the rest of the points along the control field shape. The results of this procedure are shown for two optimal control fields in Fig.~\ref{fig_sensitivity}(f) and (g), in the adiabatic [EIT-like, $\mathcal{M} = (50,1.5)$] and non-adiabatic [ATT-like, $\mathcal{M} = (50,0.01)$] regimes, respectively. Fig.~\ref{fig_sensitivity}(f) shows that EIT-like control field shapes are most sensitive to drift along the trailing edge of the control field shape, which overlaps with the signal field in time and adiabatically closes the transparency window. Importantly, this result is in agreement with the experimental determination of Ref.~\cite{guo2019high}. For ATT-like control field shapes, Fig.~\ref{fig_sensitivity}(g) shows that the most sensitive regions of the control field are those with the largest Rabi frequency. This agrees with physical intuition for the ATT protocol (see Sec.~\ref{protocol_sec} for a review), as it is these regions of the control field that have the largest effect on its net pulse area, and therefore have the largest effect on the efficiency of the atomic-polarization--to--spin-wave transfer process. The same one-at-a-time sensitivity analysis is performed for Gaussian control fields in Ref.~\cite{shinbrough2022variance}, as well as a more sophisticated and computationally expensive Sobol' analysis that probes correlations between Gaussian control field parameters.

We stress here that the sensitivity analysis performed in Ref.~\cite{shinbrough2022variance} represents only a small subset of all sensitivity calculations that can be performed, and which may be relevant and useful for quantum memory experiments. The general framework developed in Ref.~\cite{shinbrough2022variance} may also be applied to characterize the sensitivity of off-resonant memory protocols, or those protocols that make use of inhomogeneously broadened ensembles. Moreover, Ref.~\cite{shinbrough2022variance} considers only the case of `efficiency sensitivity,' or how fluctuations and drift in certain parameters affect memory efficiency. The same approach may straightforwardly be repurposed to evaluate such metrics as fidelity sensitivity, bandwidth sensitivity, or lifetime sensitivity, to name only a few. This may be the subject of future work.

\section{State of the Art}\label{SotA_sec}

\begin{figure}[H]
	\centering
	\includegraphics[width=0.82\linewidth]{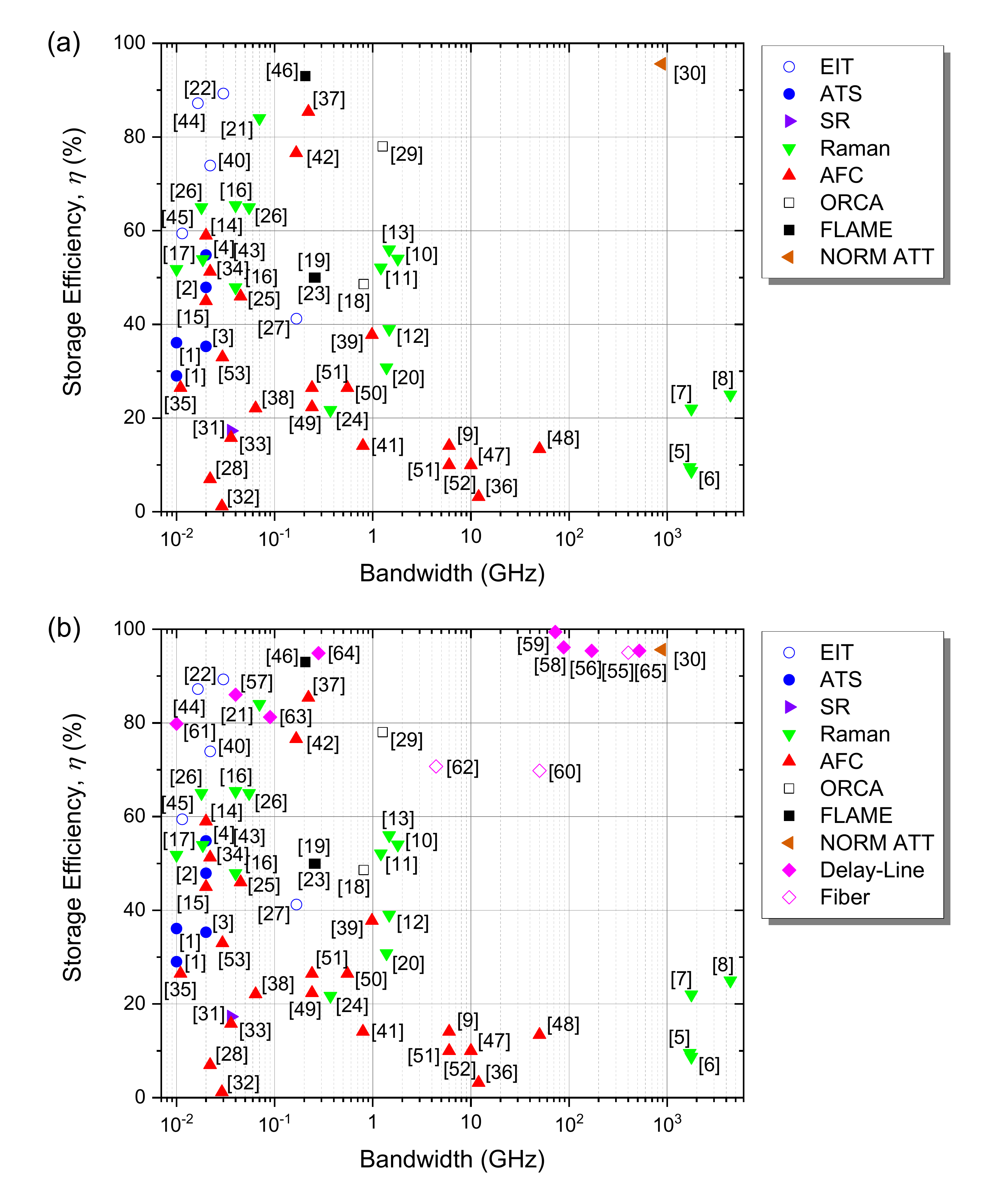}
	\caption{State of the art storage efficiencies for (a) atomic ensemble and (b) all quantum memories (including delay-based) in the broadband regime. Delay-line and fiber memories tend to have high efficiency independent of bandwidth. Citation numbers appear in Ref.~\cite{SotAfigRefs}. Adapted from Ref.~\cite{shinbrough2022broadband}.} 
	\label{fig_eff_v_bw}
\end{figure}

\clearpage

\begin{figure}[H]
	\centering
	\includegraphics[width=0.8\linewidth]{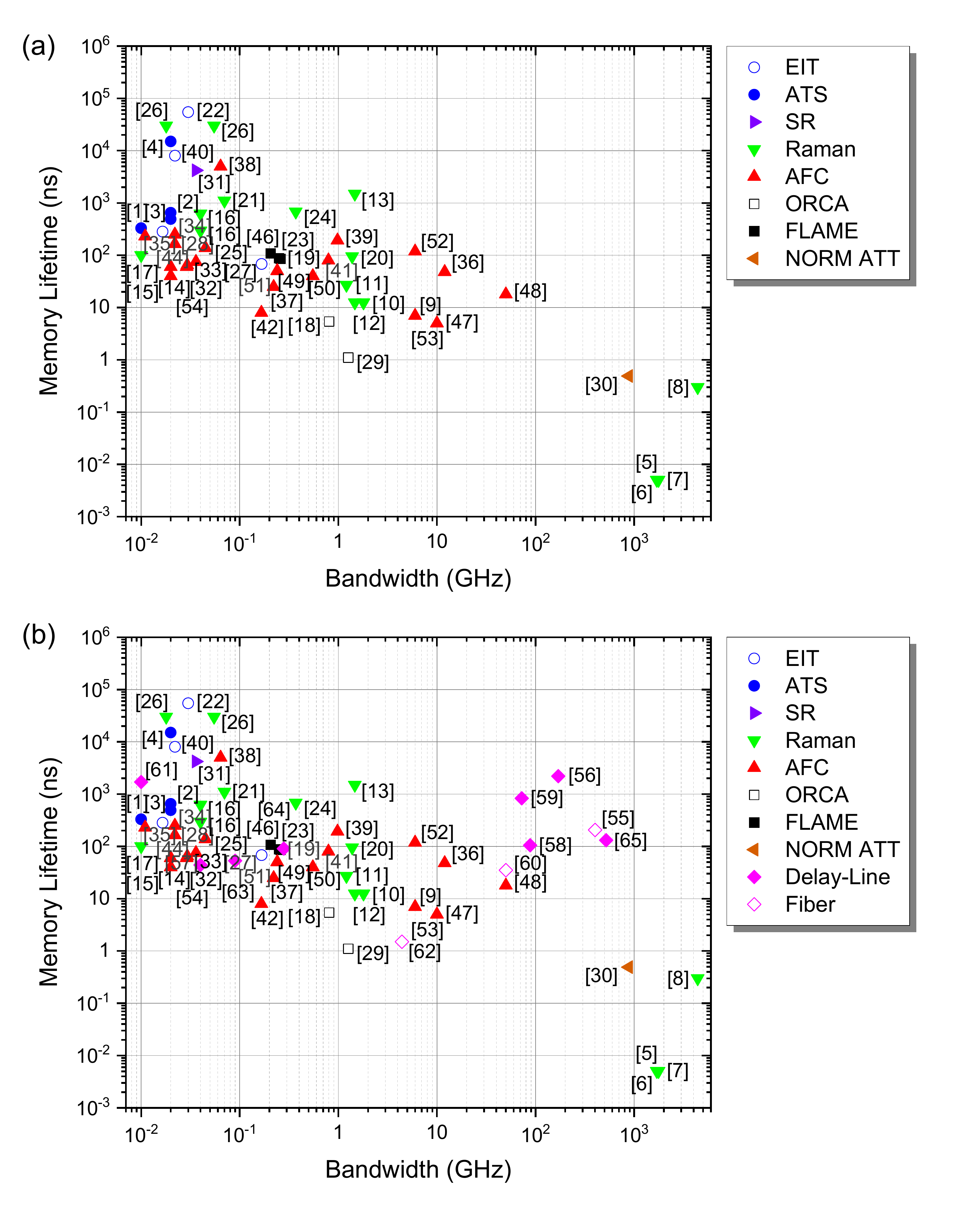}
	\caption{State of the art memory lifetimes for (a) atomic ensemble and (b) all quantum memories (including delay-based) in the broadband regime. Citation numbers appear in Ref.~\cite{SotAfigRefs}. Adapted from Ref.~\cite{shinbrough2022efficient}.}
	\label{fig_lifetime_v_bw}
\end{figure}

\clearpage

\subsection{Efficiency}\label{eff_sec} 

With the foundation for the quantum memory protocols and theory laid above, in this section we move on to discuss the state of the art for experimental quantum memory performance in the broadband regime, which we consider to be memory bandwidths greater than 10 MHz (defined as the full width at half-maximum of the signal frequency intensity distribution, see Sec.~\ref{metric_sec}). 

Figure~\ref{fig_eff_v_bw}(a) shows the state-of-the-art storage efficiencies for atomic ensemble quantum memories in the broadband regime, as of the writing of this chapter. Each memory protocol is given a different marker type. As can be seen, Raman and AFC memories are the most popular types in the broadband regime, and Raman memories are the only type used in the THz-bandwidth regime except for Ref.~[30], which most closely resembles the absorb-then-transfer protocol. Resonant, homogeneous atomic ensemble protocols (e.g., EIT, ATS, SR, FLAME) tend to be limited to bandwidths below 1 GHz, with the exception of Ref.~[30], due to linewidth-bandwidth mismatch. More details on how Ref.~[30] alleviates the linewidth-bandwidth mismatch problem can be found in Sec.~\ref{bwlw_sec}. Fig.~\ref{fig_eff_v_bw}(b) shows the performance of all quantum memories in the broadband regime, as of the writing of this chapter, including delay-line and fiber memories, which tend to have high efficiency independent of bandwidth. 

In Fig.~\ref{fig_eff_v_bw}, we show state-of-the-art storage efficiencies; total efficiency follows a similar trend, but is always significantly lower, and the trend is complicated by non-optimal phasematching and reabsorption loss during retrieval, as well as spin wave decay, which vary depending on the reference and reduce total efficiency non-uniformly across the bandwidths shown.

\subsection{Memory Lifetime}\label{lifetime_sec}

Figure~\ref{fig_lifetime_v_bw} shows the state-of-the-art memory lifetimes for atomic ensemble quantum memories [Fig.~\ref{fig_lifetime_v_bw}(a)] and all quantum memories [Fig.~\ref{fig_lifetime_v_bw}(b)] in the broadband regime. An empirical tradeoff exists between memory lifetime and bandwidth. We believe this trend primarily arises from technical considerations, rather than fundamental physical ones --- in the THz-bandwidth regime, for example, the Raman memories shown suffer from low memory lifetime not due to a fundamental aspect of the Raman protocol, but because the hardware employed (solids and molecular gases) have inherently limited storage state lifetimes. Ref.~[30] represents a unique case in this regard, where the lifetime is instead limited by atomic motion (sometimes called Doppler broadening) due to the high temperatures used, yet the inherent storage state lifetime is long, of order 0.1 seconds \cite{migdalek1990multiconfiguration}. The delay-based memories shown in Fig.~\ref{fig_lifetime_v_bw}(b) break this trend, and all tend to have nanosecond to microsecond delay times.

\subsection{Noise}\label{noise_sec}

\begin{figure}[t]
	\centering
	\includegraphics[width=0.82\linewidth]{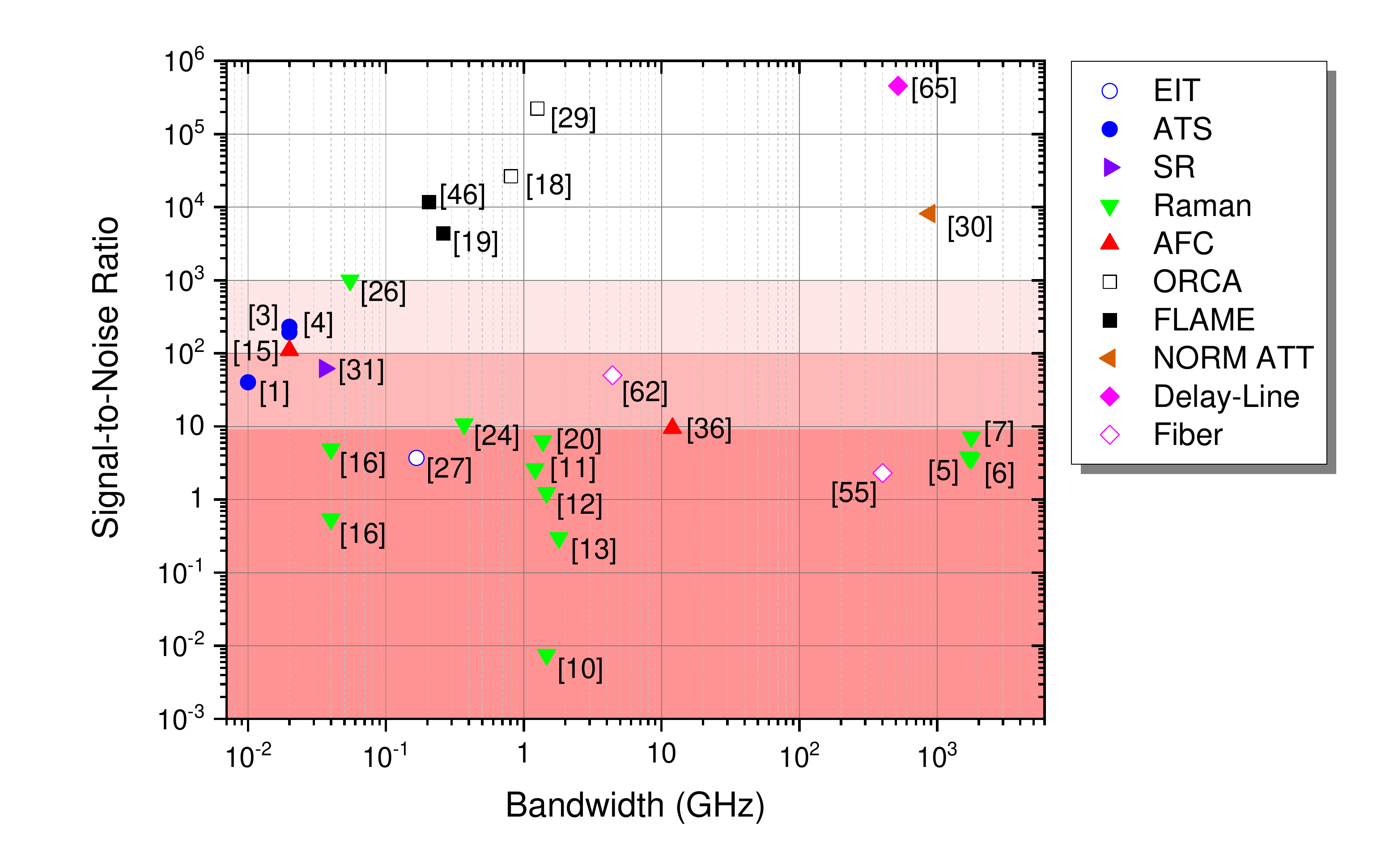}
	\caption{State-of-the-art noise performance for atomic ensemble and delay-based quantum memories in the broadband regime. Dark, medium, and light red regions correspond to the signal to noise ratios necessary for $<$90\%, $<$99\%, and $<$99.9\% single-photon fidelities. Note that many fiber and delay-line memories do not report a figure of merit for noise performance, and are therefore not plotted here, as the added noise is virtually zero; some of those presented here are fiber memories that rely on nonlinear optical interaction with strong laser light, which imparts added noise, and are not representative of all fiber or delay-line memories. Citation numbers appear in Ref.~\cite{SotAfigRefs}.}
	\label{fig_noise_v_bw}
\end{figure}

We plot the state-of-the-art noise performance for broadband atomic ensemble quantum memories in Figure~\ref{fig_noise_v_bw}. We plot the signal-to-noise ratio as defined in Sec.~\ref{metric_sec} for each broadband memory found in the literature. The dark, medium, and light red regions represent the signal to noise ratios for $<$90\%, $<$99\%, and $<$99.9\% single-photon fidelities. Across all bandwidths shown, only a few quantum memories exceed an $\mathrm{SNR}$ of 10$^3$; most memories in this region employ a ladder-type energy level system (ORCA and FLAME memories, specifically) that is in principle noise-free. Ref.~[30] is the only atomic ensemble memory in this regime that employs a $\Lambda$-type level structure, and is also in principle noise-free to first order due to a large ground-state--storage-state splitting compared to the storage-state--excited-state splitting.

\section{Conclusion}\label{conc_sec}

In this work we have comprehensively reviewed broadband quantum memory in atomic ensembles, including its motivation, the challenge posed by linewidth-bandwidth mismatch, the physical protocols applicable to atomic ensembles, the underlying theory for homogeneous and inhomogeneous systems and application of the theory for efficiency optimization and sensitivity analysis, and the current state-of-the-art performance of broadband atomic ensembles relative to delay-line and fiber memories. We hope this chapter serves as a useful guide to and reference for broadband atomic ensemble quantum memory.

\section*{Acknowledgements}

This work was funded in part by NSF grant Nos. 1640968, 1806572, 1839177, 1936321, and 2207822; and NSF Award DMR1747426. This material is based upon work supported by the U.S. Department of Energy Office of Science National Quantum Information Science Research Centers. We thank J. Gary Eden, Benjamin D. Hunt, Sehyun Park, Andrey Mironov, Thomas Reboli, Kathleen Oolman, Yujie Zhang, Dongbeom Kim, Yunkai Wang, Colin Lualdi, Tegan Loveridge, and Ujaan Purakayastha for helpful discussion, and Jim Brownfield and Ernest Northern for expert assistance machining the barium heat pipe oven.

\bibliographystyle{apsrev4-1}

\end{document}